\pdfoutput=1
\documentclass[12pt,a4paper]{article}

\usepackage{lineno}  
\usepackage{graphicx}  
\usepackage{xspace} 
\usepackage{color}
\usepackage{colortbl}
\usepackage{amsmath} 
\usepackage{ifthen} 
\graphicspath{{./figs/}} 

\newboolean{pdflatex}
\setboolean{pdflatex}{true} 
\newboolean{articletitles}
\setboolean{articletitles}{true} 

\newboolean{uprightparticles}
\setboolean{uprightparticles}{false} 

\usepackage{amssymb}
\usepackage{amsfonts}
\usepackage{upgreek} 

\usepackage[bookmarks=false]{hyperref}
\usepackage[all]{hypcap} 

\def\lhcb {\mbox{LHCb}\xspace}
\def\ux85 {\mbox{UX85}\xspace}

\ifthenelse{\boolean{uprightparticles}}%
{

 \def\Ppsi        {\ensuremath{\uppsi}\xspace}

 \def\PDelta      {\ensuremath{\Delta}\xspace}                 
 \def\PXi      {\ensuremath{\Xi}\xspace}                 
 \def\PLambda      {\ensuremath{\Lambda}\xspace}                 
 \def\PSigma      {\ensuremath{\Sigma}\xspace}                 
 \def\POmega      {\ensuremath{\Omega}\xspace}                 
 \def\PUpsilon      {\ensuremath{\Upsilon}\xspace}                 
 
 \def\PB      {\ensuremath{\mathrm{B}}\xspace}                 
                  
 \def\PD      {\ensuremath{\mathrm{D}}\xspace}

 \def\PJ      {\ensuremath{\mathrm{J}}\xspace}                 
 \def\PK      {\ensuremath{\mathrm{K}}\xspace}

 \def\Pb      {\ensuremath{\mathrm{b}}\xspace}                 
 \def\Pc      {\ensuremath{\mathrm{c}}\xspace}

 \def\Pi      {\ensuremath{\mathrm{i}}\xspace}

 \def\Ps      {\ensuremath{\mathrm{s}}\xspace}

}
{

 \def\Ppsi        {\ensuremath{\psi}\xspace}                 
                  
 \mathchardef\PDelta="7101
 \mathchardef\PXi="7104
 \mathchardef\PLambda="7103
 \mathchardef\PSigma="7106
 \mathchardef\POmega="710A
 \mathchardef\PUpsilon="7107
                  
 \def\PB      {\ensuremath{B}\xspace}                 
                  
 \def\PD      {\ensuremath{D}\xspace}

 \def\PJ      {\ensuremath{J}\xspace}                 
 \def\PK      {\ensuremath{K}\xspace}

 \def\Pb      {\ensuremath{b}\xspace}                 
 \def\Pc      {\ensuremath{c}\xspace}

 \def\Pi      {\ensuremath{i}\xspace}

 \def\Ps      {\ensuremath{s}\xspace}

}

\def\squark    {\ensuremath{\Ps}\xspace}

\def\cquark    {\ensuremath{\Pc}\xspace}

\def\bquark    {\ensuremath{\Pb}\xspace}

\def\kaon  {\ensuremath{\PK}\xspace}
  \def\Kbar  {\kern 0.2em\overline{\kern -0.2em \PK}{}\xspace}

\def\Kz    {\ensuremath{\kaon^0}\xspace}
\def\Kzb   {\ensuremath{\Kbar^0}\xspace}
\def\KzKzb {\ensuremath{\Kz \kern -0.16em \Kzb}\xspace}
\def\Kp    {\ensuremath{\kaon^+}\xspace}
\def\Km    {\ensuremath{\kaon^-}\xspace}

\def\KpKm  {\ensuremath{\Kp \kern -0.16em \Km}\xspace}

  \def\Dbar    {\kern 0.2em\overline{\kern -0.2em \PD}{}\xspace}
\def\D       {\ensuremath{\PD}\xspace}

\def\Dz      {\ensuremath{\D^0}\xspace}
\def\Dzb     {\ensuremath{\Dbar^0}\xspace}
\def\DzDzb   {\ensuremath{\Dz {\kern -0.16em \Dzb}}\xspace}
\def\Dp      {\ensuremath{\D^+}\xspace}
\def\Dm      {\ensuremath{\D^-}\xspace}

\def\DpDm    {\ensuremath{\Dp {\kern -0.16em \Dm}}\xspace}

\def\B       {\ensuremath{\PB}\xspace}
  \def\Bbar    {\kern 0.18em\overline{\kern -0.18em \PB}{}\xspace}

\def\Bz      {\ensuremath{\B^0}\xspace}
\def\Bzb     {\ensuremath{\Bbar^0}\xspace}

\def\Bs      {\ensuremath{\B^0_\squark}\xspace}
\def\Bsb     {\ensuremath{\Bbar^0_\squark}\xspace}
\def\Bdb     {\ensuremath{\Bbar^0}\xspace}

\def\jpsi     {\ensuremath{{\PJ\mskip -3mu/\mskip -2mu\Ppsi\mskip 2mu}}\xspace}

  \def\Y#1S{\ensuremath{\PUpsilon{(#1S)}}\xspace}

\def\Lbar {\ensuremath{\kern 0.1em\overline{\kern -0.1em\Lambda\kern -0.05em}\kern 0.05em{}}\xspace}

\def\to                 {\ensuremath{\rightarrow}\xspace}

\def\CP                {\ensuremath{C\!P}\xspace}

\def\AT#1     {\ensuremath{A_{\mathrm{T}}^{#1}}\xspace}           

\def\C#1      {\ensuremath{\mathcal{C}_{#1}}\xspace}                       
\def\Cp#1     {\ensuremath{\mathcal{C}_{#1}^{'}}\xspace}                    
\def\Ceff#1   {\ensuremath{\mathcal{C}_{#1}^{\mathrm{(eff)}}}\xspace}        
\def\Cpeff#1  {\ensuremath{\mathcal{C}_{#1}^{'\mathrm{(eff)}}}\xspace}       
\def\Ope#1    {\ensuremath{\mathcal{O}_{#1}}\xspace}                       
\def\Opep#1   {\ensuremath{\mathcal{O}_{#1}^{'}}\xspace}                    

\newcommand{\tev}{\ensuremath{\mathrm{\,Te\kern -0.1em V}}\xspace}
\newcommand{\gev}{\ensuremath{\mathrm{\,Ge\kern -0.1em V}}\xspace}
\newcommand{\mev}{\ensuremath{\mathrm{\,Me\kern -0.1em V}}\xspace}
\newcommand{\kev}{\ensuremath{\mathrm{\,ke\kern -0.1em V}}\xspace}
\newcommand{\ev}{\ensuremath{\mathrm{\,e\kern -0.1em V}}\xspace}
\newcommand{\gevc}{\ensuremath{{\mathrm{\,Ge\kern -0.1em V\!/}c}}\xspace}
\newcommand{\mevc}{\ensuremath{{\mathrm{\,Me\kern -0.1em V\!/}c}}\xspace}
\newcommand{\gevcc}{\ensuremath{{\mathrm{\,Ge\kern -0.1em V\!/}c^2}}\xspace}
\newcommand{\gevgevcccc}{\ensuremath{{\mathrm{\,Ge\kern -0.1em V^2\!/}c^4}}\xspace}
\newcommand{\mevcc}{\ensuremath{{\mathrm{\,Me\kern -0.1em V\!/}c^2}}\xspace}

\def\mum  {\ensuremath{\,\upmu\rm m}\xspace}

\def\gsim{{~\raise.15em\hbox{$>$}\kern-.85em
          \lower.35em\hbox{$\sim$}~}\xspace}
\def\lsim{{~\raise.15em\hbox{$<$}\kern-.85em
          \lower.35em\hbox{$\sim$}~}\xspace}

\def\pt         {\mbox{$p_{\rm T}$}\xspace}

\def\pythia     {\mbox{\textsc{Pythia}}\xspace}

\def\gauss      {\mbox{\textsc{Gauss}}\xspace}

\def\tell1  {TELL1\xspace}
\def\ukl1   {UKL1\xspace}

\usepackage{cite}
\usepackage{mciteplus}

\begin{document}
\renewcommand{\thefootnote}{\fnsymbol{footnote}}
\setcounter{footnote}{1}



\begin{titlepage}
\pagenumbering{roman}
\belowpdfbookmark{Title page}{title}

\pagenumbering{roman}
\vspace*{-1.5cm}
\centerline{\large EUROPEAN ORGANIZATION FOR NUCLEAR RESEARCH (CERN)}
\vspace*{1.5cm}
\hspace*{-5mm}\begin{tabular*}{16cm}{lc@{\extracolsep{\fill}}r}
 & & CERN-PH-EP-2012-111\\
 & & LHCb-PAPER-2012-005\\  
 & & April 25, 2012 \\ 
 & & \\
\hline
\end{tabular*}

\vspace*{2.5cm}

{\bf\boldmath\huge
\begin{center}
Analysis of the resonant components in $\Bsb\rightarrow J/\psi \pi^+\pi^-$
\end{center}
}

\vspace*{2.0cm}
\begin{center}
\normalsize {
The LHCb collaboration\footnote{Authors are listed on the following pages.}
}
\end{center}

\vspace*{.5cm}

\begin{abstract}
  \noindent

The decay $\Bsb\rightarrow J/\psi \pi^+\pi^-$ can be exploited to study \CP violation. A detailed understanding of its structure is imperative in order to optimize its usefulness. An analysis of this three-body final state is performed using a 1.0\,{fb}$^{-1}$ sample of data produced in 7 TeV $pp$ collisions at the LHC and collected by the LHCb experiment.
A modified Dalitz plot analysis of the final state is performed using both the invariant mass spectra and the decay angular distributions. The $\pi^+\pi^-$ system is shown to be dominantly in an S-wave state, and the \CP-odd fraction in this \Bsb decay is shown to be greater than 0.977 at 95\% confidence level. In addition, we report the first measurement of the $\jpsi\pi^+\pi^-$ branching fraction relative to $J/\psi\phi$ of $(19.79\pm 0.47 \pm 0.52)$\%.
 
\end{abstract}

\vspace*{2.5cm}

\hspace*{6mm}Submitted to Physics Review D\\
\newpage
\begin{center}
{\bf LHCb collaboration}\\
\end{center}

\begin{flushleft}
R.~Aaij$^{38}$, 
C.~Abellan~Beteta$^{33,n}$, 
B.~Adeva$^{34}$, 
M.~Adinolfi$^{43}$, 
C.~Adrover$^{6}$, 
A.~Affolder$^{49}$, 
Z.~Ajaltouni$^{5}$, 
J.~Albrecht$^{35}$, 
F.~Alessio$^{35}$, 
M.~Alexander$^{48}$
% ...

{\footnotesize \it
$ ^{1}$Centro Brasileiro de Pesquisas F\'{i}sicas (CBPF), Rio de Janeiro, Brazil\\
% ...
$ ^{5}$Clermont Universit\'{e}, Universit\'{e} Blaise Pascal, CNRS/IN2P3, LPC, Clermont-Ferrand, France\\
$ ^{6}$CPPM, Aix-Marseille Universit\'{e}, CNRS/IN2P3, Marseille, France\\
% ...
$ ^{33}$Universitat de Barcelona, Barcelona, Spain\\
$ ^{34}$Universidad de Santiago de Compostela, Santiago de Compostela, Spain\\
$ ^{35}$European Organization for Nuclear Research (CERN), Geneva, Switzerland\\
% ...
$ ^{38}$Nikhef National Institute for Subatomic Physics, Amsterdam, The Netherlands\\
% ...
$ ^{43}$H.H. Wills Physics Laboratory, University of Bristol, Bristol, United Kingdom\\
% ...
$ ^{48}$School of Physics and Astronomy, University of Glasgow, Glasgow, United Kingdom\\
$ ^{49}$Oliver Lodge Laboratory, University of Liverpool, Liverpool, United Kingdom\\
% ...
$ ^{n}$LIFAELS, La Salle, Universitat Ramon Llull, Barcelona, Spain\\
% ...
}
\end{flushleft}

\end{titlepage}

\renewcommand{\thefootnote}{\arabic{footnote}}
\setcounter{footnote}{0}

\pagestyle{empty}  



\setcounter{page}{2}
\mbox{~}




\pagestyle{plain} 
\setcounter{page}{1}
\pagenumbering{arabic}


%
\section{Introduction}
\label{sec:Introduction}

\begin{figure}[h]
\vskip -.4cm
\begin{center}
\includegraphics[width=3in]{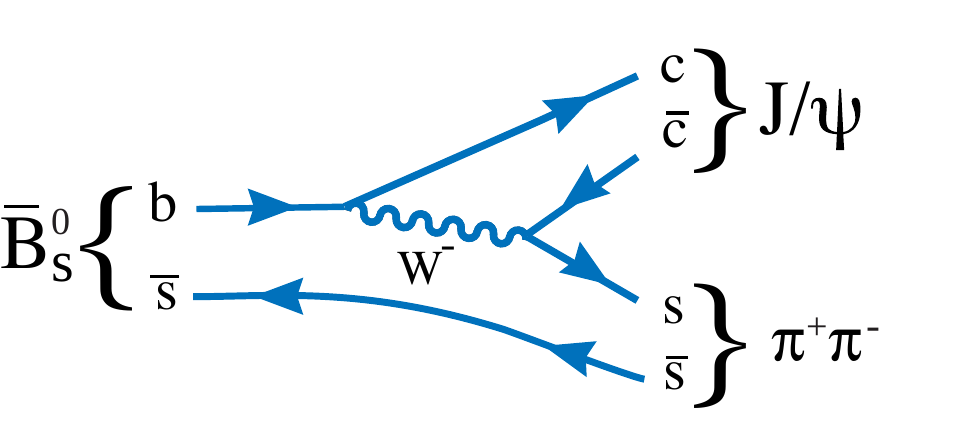}
\end{center}\label{feyn1}
\vskip -.5cm
\caption{Leading order diagram for $\Bsb$ decays into $J/\psi \pi^+\pi^-$.}
\end{figure}

Measurement of mixing-induced \CP violation in $\Bsb$ decays is of prime importance in probing physics beyond the Standard Model. 
Final states that are \CP eigenstates with large rates and high detection efficiencies are very useful for such studies. The $\Bsb\rightarrow J/\psi f_0(980)$, $f_0(980)\to\pi^+\pi^-$ decay mode, a \CP-odd eigenstate, was discovered by the LHCb collaboration \cite{Aaij:2011fx} and subsequently confirmed by several experiments \cite{Li:2011pg,*Abazov:2011hv,*Aaltonen:2011nk}. As we use the $J/\psi\to\mu^+\mu^-$ decay, the final state has four charged tracks, and has high detection efficiency.  LHCb has used this mode to measure the  \CP violating phase $\phi_s$ \cite{LHCb:2011ab}, which complements measurements in the $J/\psi\phi$ final state \cite{LHCb:2011aa,CDF:2011af,*Abazov:2011ry}. 
It is possible that a larger $\pi^+\pi^-$ mass range could also be used for such studies. Therefore, to fully exploit the $J/\psi \pi^+\pi^-$ final state for measuring \CP violation, it is important to determine its resonant and \CP content.
The tree-level Feynman diagram for the process is shown in Fig.~\ref{feyn1}.

In this paper the $J/\psi\pi^+$ and $\pi^+\pi^-$ mass spectra, and decay angular distributions are used to study the resonant and non-resonant structures. This differs from a classical ``Dalitz plot" analysis \cite{Dalitz:1953cp} because one of the particles in the final state, the $J/\psi$, has spin-1 and its three decay amplitudes must be considered. We first show that there are no evident structures in the $J/\psi\pi^+$ invariant mass, and then model the $\pi^+\pi^-$ invariant mass with a series of resonant and non-resonant amplitudes. The data are then fitted with the coherent sum of these amplitudes. We report on the resonant structure and the \CP content of the final state.

\section{Data sample and analysis requirements}

The data sample contains 1.0\,{fb}$^{-1}$ of integrated luminosity collected with the \lhcb detector  \cite{LHCb-det} using $pp$ collisions at a center-of-mass energy of 7 TeV. 
The detector is a single-arm forward
spectrometer covering the pseudorapidity range $2<\eta <5$, designed
for the study of particles containing \bquark or \cquark quarks. Components include a high precision tracking system consisting of a
silicon-strip vertex detector surrounding the $pp$ interaction region,
a large-area silicon-strip detector located upstream of a dipole
magnet with a bending power of about $4{\rm\,Tm}$, and three stations
of silicon-strip detectors and straw drift-tubes placed
downstream. The combined tracking system has a momentum resolution
$\Delta p/p$ that varies from 0.4\% at 5\gev to 0.6\% at 100\gev (we work in units where $c=1$), 
and an impact parameter resolution of 20\mum for tracks with large
transverse momentum with respect to the proton beam direction. Charged hadrons are identified using two
ring-imaging Cherenkov (RICH) detectors. Photon, electron and hadron
candidates are identified by a calorimeter system consisting of
scintillating-pad and pre-shower detectors, an electromagnetic
calorimeter and a hadronic calorimeter. Muons are identified by a muon
system composed of alternating layers of iron and multiwire
proportional chambers. The trigger consists of a hardware stage, based
on information from the calorimeter and muon systems, followed by a
software stage which applies a full event reconstruction.

 Events selected for this analysis are triggered by a $J/\psi\to\mu^+\mu^-$ decay.
Muon candidates are selected at the hardware level using their penetration through iron and detection in a series of tracking chambers. They are also required in the software level to be consistent with coming from the decay of a $\Bsb$ meson into a $J/\psi$. Only \jpsi decays that are triggered on are used.

\section{Selection requirements}
\label{sec:2}
The selection requirements discussed here are imposed to isolate $\Bsb$ candidates with high signal yield and minimum background. This is accomplished by first selecting candidate $J/\psi\to\mu^+\mu^-$ decays, selecting a pair of pion candidates of opposite charge, and then testing if all four tracks form a common decay vertex.
To be considered
a $J/\psi\to\mu^+\mu^-$ candidate particles of opposite charge are required to have  transverse momentum, $p_{\rm T}$, greater than 500\,MeV, be identified as muons, and 
form a vertex with fit $\chi^2$ per number of degrees of freedom (ndf) less than 11. After applying these requirements, there is a large \jpsi signal over a small background \cite{Aaij:2011fx}. Only candidates with dimuon invariant mass between $-$48~MeV to +43 MeV relative to the observed $J/\psi$ mass peak are selected. The requirement is asymmetric because of final state electromagnetic radiation. The two muons subsequently are kinematically constrained to the known $J/\psi$ mass~\cite{PDG}. 

Pion and kaon candidates are positively identified using the RICH system. Cherenkov photons are matched to charged tracks, the emission angles of the photons compared with those expected if the particle is an electron, pion, kaon or proton, and a likelihood is then computed.  The particle identification is done by using the logarithm of the likelihood ratio comparing two particle hypotheses (DLL).  For pion selection we require DLL$(\pi-K)>-10$.

Candidate $\pi^+\pi^-$ combinations are selected if each particle is inconsistent with having been produced at the primary vertex. This is done by use of the impact parameter (IP) defined as the minimum distance of approach of the track with respect to the primary vertex. We require that the $\chi^2$ formed by using the hypothesis that the IP is zero be greater than 9 for each track. Furthermore,  each pion candidate must have  $p_{\rm T}>250$ MeV and the scalar sum of the two pion candidate momentum,  $p_{\rm T}(\pi^+)+p_{\rm T}(\pi^-)$,  must be greater than 900\,MeV.
To select $\Bsb$ candidates we further require that the
two pion candidates form a vertex with a $\chi^2< 10$, that they form a candidate $\Bsb$ vertex with the $J/\psi$ where the vertex fit $\chi^2$/ndf $<5$, that this vertex is greater than $1.5$\,mm from the primary vertex and  the angle between the  \Bsb momentum vector and the vector from the primary vertex to the \Bsb vertex must be less than 11.8\,mrad
 
\begin{figure}[t!t!]
\begin{center}
    \includegraphics[width=6 in]{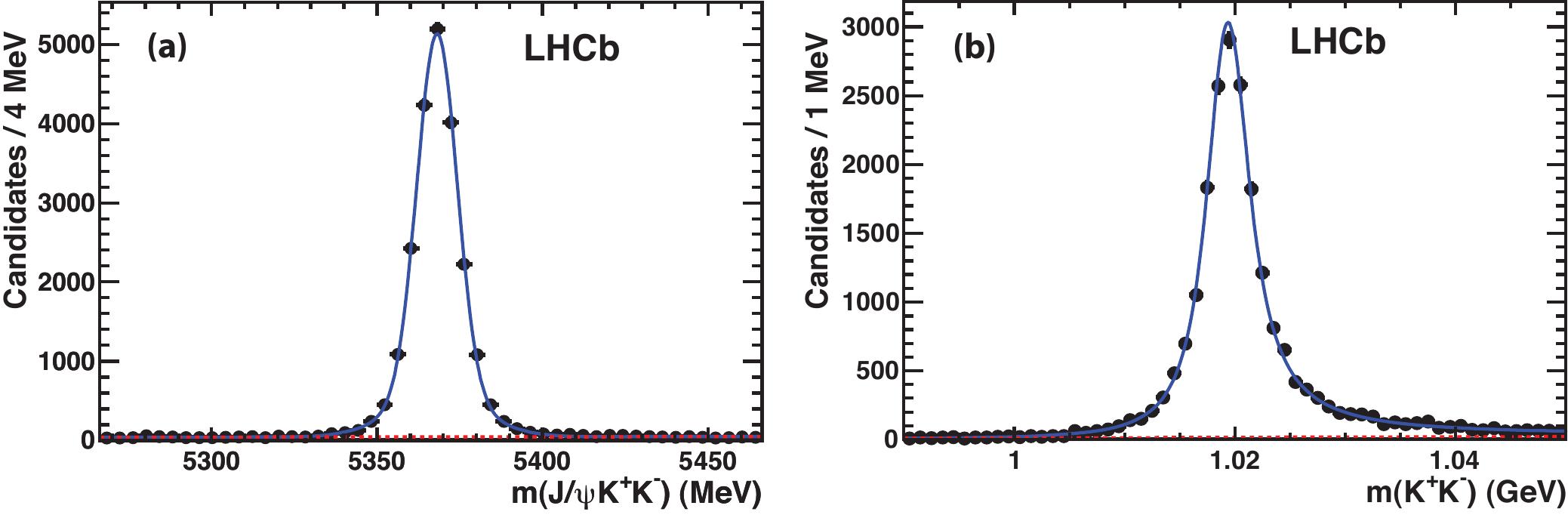}
    \caption{ (a) Invariant mass spectrum of $J/\psi K^+K^-$ for candidates with $m(K^+K^-)<1050$ MeV. The data has been fitted with a double-Gaussian signal and linear background functions shown as a dashed line. The solid curve shows the sum. (b) Background subtracted invariant mass spectrum of $K^+K^-$ for events with $m(K^+K^-)<1050$ MeV. The dashed line (barely visible along the $x$-axis) shows the S-wave contribution and  the solid curve is the sum of the S-wave and a P-wave Breit-Wigner functions, fitted to the data.}
\end{center}\label{JpsiKK}
\end{figure}

We use the decay $\Bsb\to J/\psi\phi$, $\phi\to K^+K^-$ as a normalization and control channel in this paper.
 The selection criteria are identical to the ones used for $J/\psi \pi^+\pi^-$ except for the particle identification requirement. Kaon candidates are selected requiring that DLL($K-\pi)>0$. 
 Figure~\ref{JpsiKK}(a) shows the $J/\psi K^+ K^-$ mass for all events with  $m(K^+K^-)<1050$ MeV. The $K^+K^-$ combination is not, however, pure $\phi$ due to the presence of an S-wave contribution~\cite{Stone:2008ak}. 
We determine the $\phi$ yield by fitting the data to a relativistic P-wave Breit-Wigner function that is convolved with a Gaussian function to account for the experimental mass resolution and a straight line for the S-wave. We use the ${_SPlot}$ method to subtract the background \cite{Pivk:2004ty}. This involves fitting the $J/\psi K^+K^-$ mass spectrum, determining the signal and background weights
and then plotting the resulting weighted mass spectrum, shown
in Fig.~\ref{JpsiKK}(b).  There is a large peak at the $\phi$ meson mass with a small S-wave component.
The mass fit gives 20,934$\pm 150$ events of which $(95.5\pm 0.3)$\% are $\phi$ and the remainder is the S-wave contribution. 

\begin{figure}[t]
\begin{center}
\includegraphics[scale=0.7]{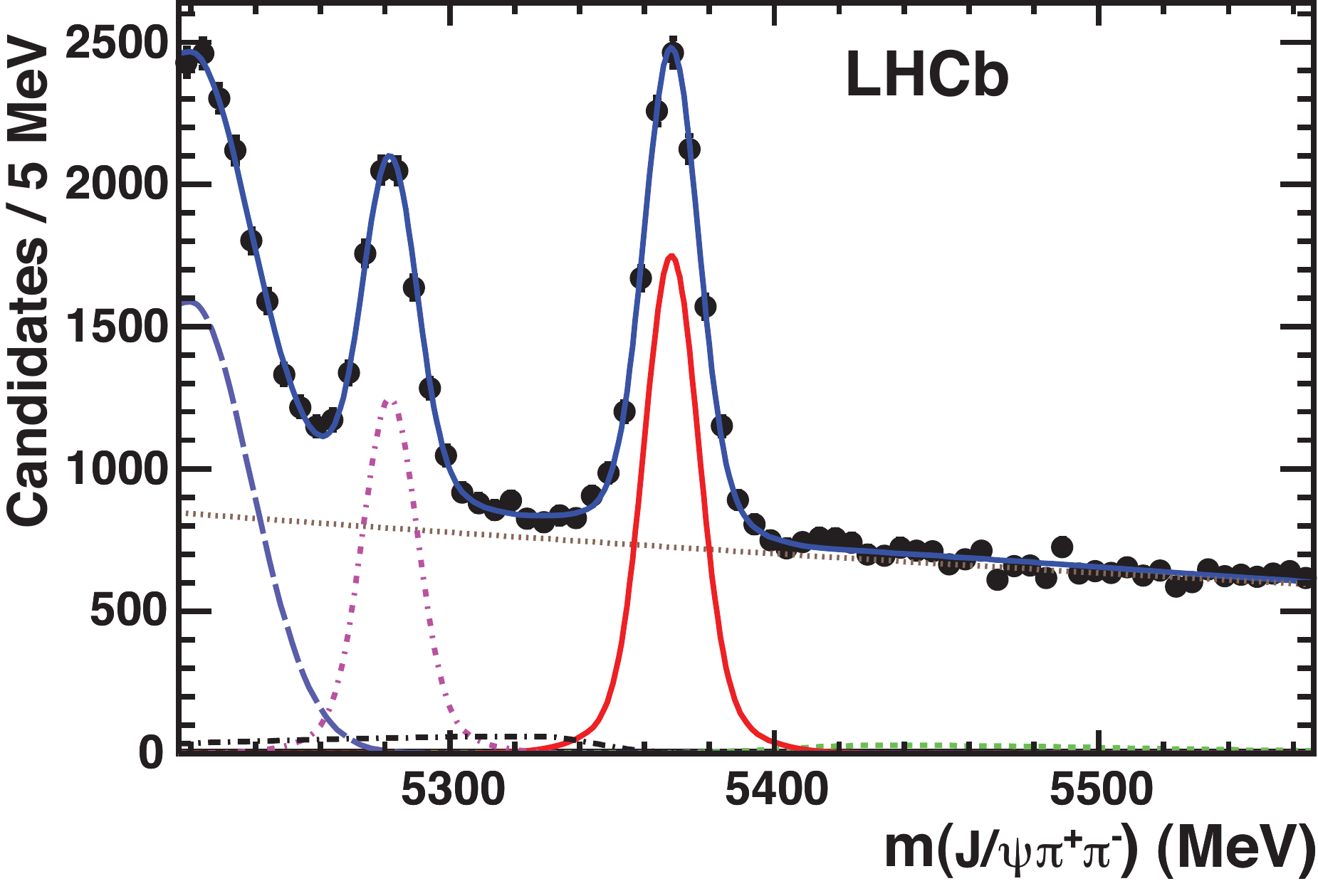}
\end{center}\label{fitmass}
\vskip -1cm
\caption{Invariant mass of $J/\psi \pi^+\pi^-$ candidate combinations. The data have been fitted with double-Gaussian signal and several background functions. The (red) solid line shows the $\Bsb$ signal, the (brown) dotted line shows the combinatorial background, the (green) short-dashed shows the $B^-$ background, the  (purple) dot-dashed is $\Bzb\rightarrow J/\psi \pi^+\pi^-$, the (black) dot-long dashed is the sum of $\Bsb\rightarrow J/\psi\eta'$ and $\Bsb\rightarrow J/\psi\phi$ when  $\phi\to\pi^+\pi^-\pi^0$ backgrounds, the (light blue) long-dashed is the $\Bdb\rightarrow J/\psi K^- \pi^+$ reflection, and the (blue) solid line is the total.}
\end{figure}

The invariant mass of the selected $J/\psi\pi^+\pi^-$ combinations, where the dimuon candidate pair is constrained to have the $J/\psi$ mass,
 is shown in Fig.~\ref{fitmass}.  There is a large peak at the $\Bsb$ mass and a smaller one at the $\Bzb$ mass on top of a background. A double-Gaussian function is used to fit the signal, the core Gaussian mean and width are allowed to vary, and the fraction and width ratio for the second Gaussian are fixed to that obtained in the fit of $\Bsb\rightarrow J/\psi \phi$. Other components in the fit model take into account contributions from  $B^-\rightarrow J/\psi K^-(\pi^-)$, $\Bsb\rightarrow J/\psi\eta' ,\eta'\rightarrow \rho \gamma$,
$\Bsb\rightarrow J/\psi\phi ,\phi\rightarrow \pi^+\pi^-\pi^0$, $\Bdb\rightarrow J/\psi \pi^+\pi^-$ backgrounds and a $\Bdb\rightarrow J/\psi K^- \pi^+$ reflection. Here and elsewhere charged conjugated modes are used when appropriate. The shape of the $\Bdb\rightarrow J/\psi \pi^+\pi^-$ signal is taken to be the same as that of the \Bsb. The exponential  combinatorial background shape is taken from wrong-sign combinations, that are the sum of $\pi^+\pi^+$ and $\pi^-\pi^-$ candidates. 
The shapes of the other components are taken from the Monte Carlo simulation with their normalizations allowed to vary (see Sect.~\ref{sec:mc}). The mass fit gives $7598\pm120$ signal and $5825\pm54$ background candidates within $\pm20$ MeV of the $\Bsb$ mass peak.

\section{Analysis formalism}\label{Formalism}

\begin{figure}[t]
\begin{center}
\includegraphics[width=4.5in]{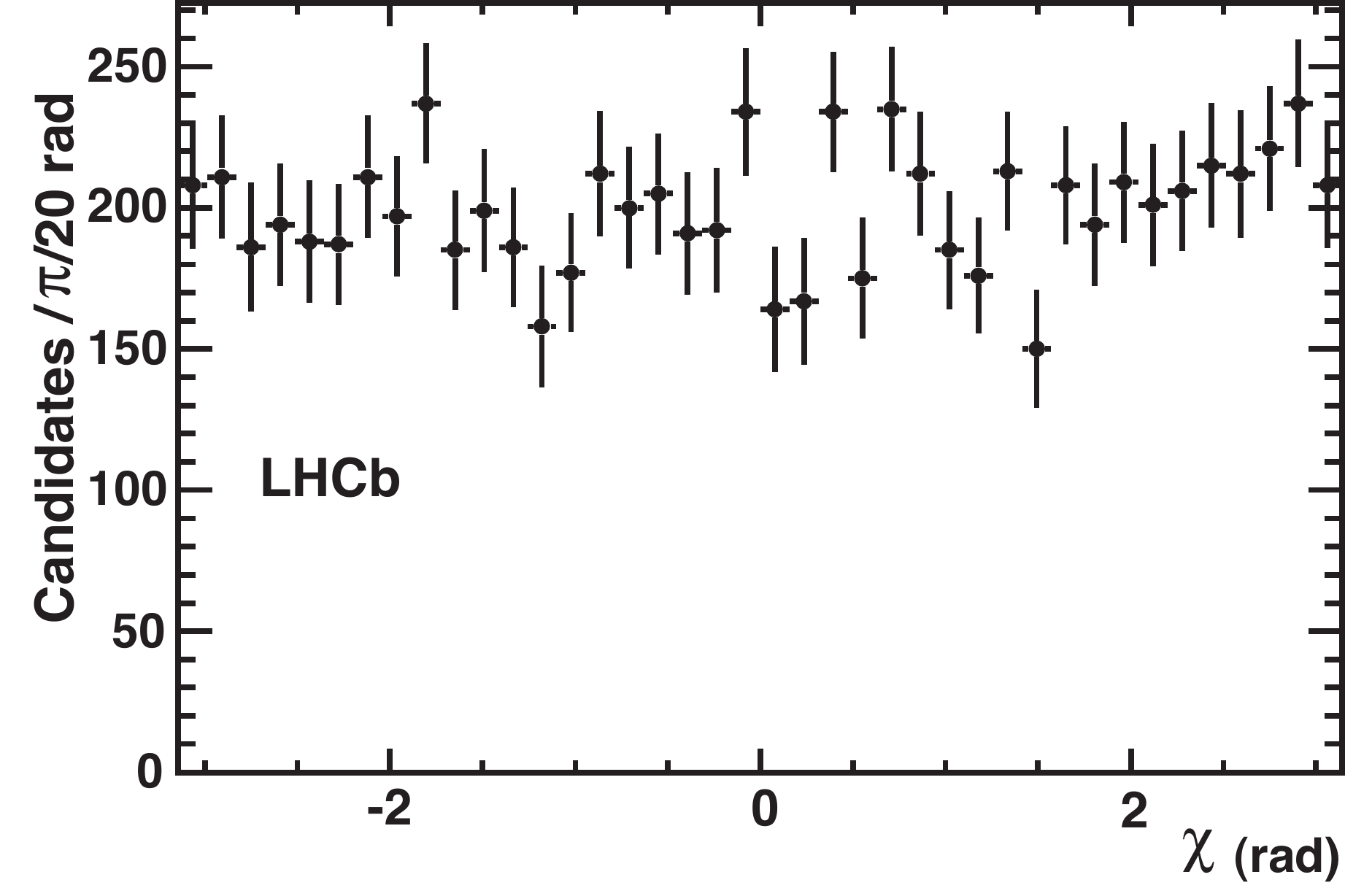}
\end{center}\label{bkgsubchi}
\vskip -5mm
\caption{Background subtracted $\chi$ distribution from $\Bsb\rightarrow J/\psi \pi^+\pi^-$ candidates. }
\end{figure}

The decay of $\Bsb\rightarrow J/\psi \pi^+\pi^-$ with the $J/\psi\rightarrow \mu^+\mu^-$ can be described by four variables. These are taken to be the invariant mass squared of $J/\psi \pi^+$ ($s_{12}\equiv m^2(J/\psi \pi^+)$), the invariant mass squared of $\pi^+\pi^-$ ($s_{23}\equiv m^2(\pi^+\pi^-)$), the $J/\psi$ helicity angle ($\theta_{J/\psi}$), which is the angle of the $\mu^+$ in the $J/\psi$ rest frame with respect to the  $\jpsi$ direction in the  $\Bsb$ rest frame,
and the angle between the $J/\psi$ and $\pi^+\pi^-$ decay planes ($\chi$) in the \Bsb rest frame. To improve the resolution of these variables we perform a kinematic fit constraining the $\Bsb$ and $J/\psi$ masses to their PDG mass values \cite{PDG}, and recompute the final state momenta.  To simplify the probability density function (PDF), we analyze the decay process after integrating over $\chi$, that eliminates several interference terms.
The $\chi$ distribution is shown in Fig.~\ref{bkgsubchi} after background subtraction using wrong-sign events. The distribution has little structure, and thus the $\chi$ acceptance can be integrated over without biasing the other variables.

\subsection{\boldmath The decay model for $\Bsb\rightarrow J/\psi \pi^+\pi^-$ }
 
One of the main challenges in performing a Dalitz plot angular analysis is to construct a realistic probability density function (PDF), where both the kinematic and dynamical properties are modeled accurately. The overall PDF given by the sum of signal, $S$, and background, $B$, functions is 
\begin{equation}
\label{eq:F}
F(s_{12}, s_{23}, \theta_{J/\psi})=\frac{f_{\rm sig}}{{\cal{N}}_{\rm sig}}\varepsilon(s_{12}, s_{23}, \theta_{J/\psi}) S(s_{12}, s_{23}, \theta_{J/\psi})+\frac{(1-f_{\rm sig})}{{\cal{N}}_{\rm bkg}} B(s_{12}, s_{23},  \theta_{J/\psi}),
\end{equation}
where  $f_{\rm sig}$ is the fraction of the signal in the fitted region and $\varepsilon$ is the detection efficiency. The normalization factors are given by
\begin{eqnarray}
{\cal{N}}_{\rm sig}&=&\int \! \varepsilon(s_{12}, s_{23}, \theta_{J/\psi}) S(s_{12}, s_{23}, \theta_{J/\psi}) \, 
ds_{12}ds_{23}d\cos\theta_{J/\psi},\nonumber\\
{\cal{N}}_{\rm bkg}&=&\int \!B(s_{12}, s_{23}, \theta_{J/\psi}) \, 
ds_{12}ds_{23}d\cos\theta_{J/\psi}.
\end{eqnarray}
In this analysis we apply a formalism similar to that used in Belle's analysis of $\Bzb\rightarrow K^-\pi^+\chi_{c1}$ decays \cite{Mizuk:2008me}.

\begin{figure}[h]
\begin{center}
\includegraphics[scale=0.5]{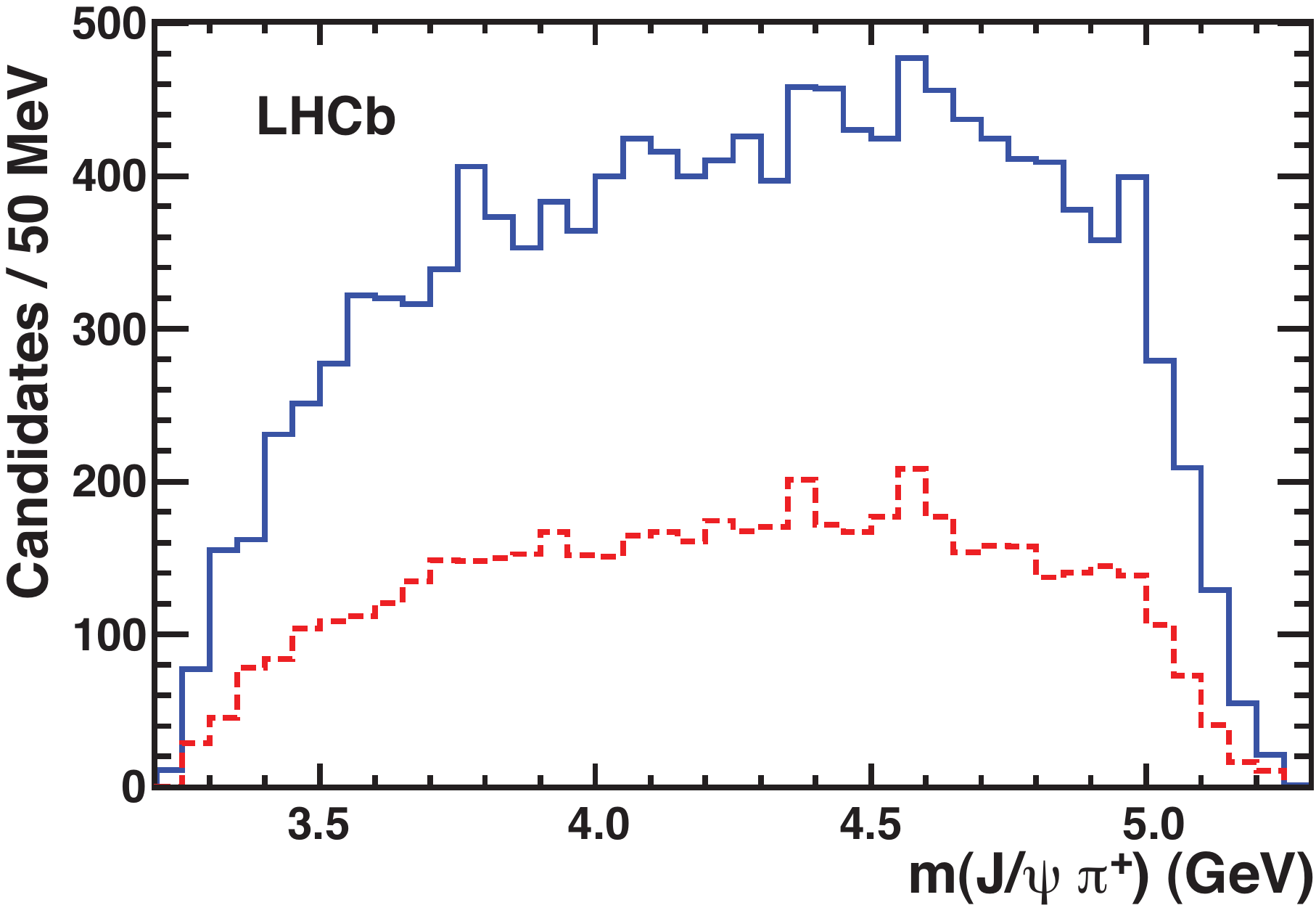}
\caption{Distribution of $m(J/\psi \pi^+)$ for $\Bsb\to J/\psi \pi^+\pi^-$ candidate decays within $\pm20$ MeV of $\Bsb$ mass shown with the (blue) solid line; $m(J/\psi \pi^+)$ for wrong-sign $J/\psi \pi^+\pi^+$ combinations is shown with the (red) dashed line, as an estimate of the background.}
\end{center}
\label{m-jpsipi}
\end{figure}

\begin{figure}[h]
\begin{center}
\includegraphics[width=4.5 in]{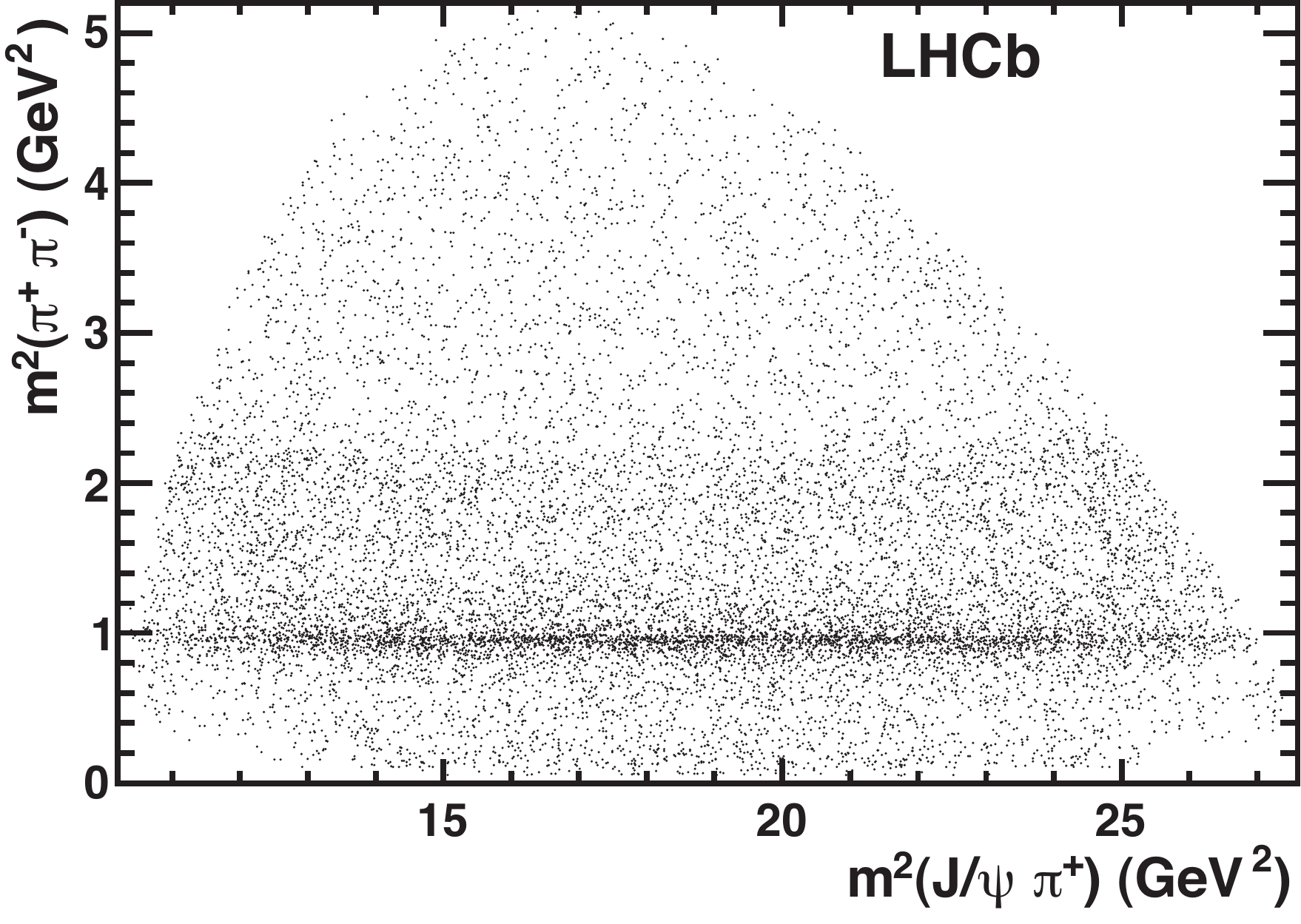}
\caption{Distribution of $s_{23}\equiv m^2(\pi^+\pi^-)$ versus $s_{12}\equiv m^2(J/\psi\pi^+)$ for $\Bsb$ candidate decays within $\pm20$ MeV of $\Bsb$ mass.}
\end{center}
\label{dalitz-1}
\end{figure}

To investigate if there are visible exotic structures in the $J/\psi\pi^+$ system as claimed in similar decays \cite{Z4430},
 we examine the $J/\psi \pi^+$  mass distribution shown in Fig.~\ref{m-jpsipi}. No resonant effects are evident.
Examination of the event distribution for $m^2(\pi^+\pi^-)$ versus $m^2(J/\psi \pi^+)$ in Fig.~\ref{dalitz-1} shows
obvious structure in $m^2(\pi^+\pi^-)$ that we wish to understand.

\subsubsection{The signal function}
 
The signal function is taken to be the sum over resonant states that can decay into  $\pi^+\pi^-$, plus a possible non-resonant S-wave contribution
\begin{equation}
S(s_{12}, s_{23}, \theta_{J/\psi})=\sum_{\lambda=0,\pm1}\left|\sum_{i}a^{R_i}_{\lambda}e^{i\phi^{R_i}_{\lambda}}
\mathcal{A}_{\lambda}^{R_i}(s_{12}, s_{23}, \theta_{J/\psi})\right|^2, \label{amplitude-eq}
\end{equation} 
where $\mathcal{A}_{\lambda}^{R_i}(s_{12}, s_{23}, \theta_{J/\psi})$ is the amplitude of the decay via an intermediate resonance $R_i$ with helicity $\lambda$. Each $R_i$ has an associated amplitude strength $a_{\lambda}^{R_i}$ for each helicity state $\lambda$ and a phase $\phi_{\lambda}^{R_i}$. The amplitudes are defined as
\begin{equation}
\mathcal{A}_{\lambda}^R(s_{12},s_{23}, \theta_{J/\psi})= F_B^{(L_B)}\; A_R(s_{23})\;F_R^{(L_R)}\; T_{\lambda} 
 \Big(\frac{P_B}{m_B}\Big )^{L_B}\; \Big( \frac{P_R}{\sqrt{s_{23}}}\Big )^{L_R}\; \Theta_{\lambda}(\theta_{J/\psi}),
\end{equation}
where $P_B$ is the \jpsi   momentum in the $\Bsb$ rest frame and $P_R$ is the momentum of
either of the two pions in the dipion rest frame, $m_{B}$ is the $\Bsb$ mass, $F_B^{(L_B)}$ and $F_R^{(L_R)}$ are the $\Bsb$ meson and $R_i$ resonance decay form factors, $L_B$ is the orbital angular momentum between the $J/\psi$ and $\pi^+\pi^-$ system, and $L_R$ the orbital angular momentum in the $\pi^+\pi^-$ decay, and thus is the same as the spin of the $\pi^+\pi^-$. Since the parent $\Bsb$ has spin-0 and the $J/\psi$ is a vector, when the $\pi^+\pi^-$ system forms a spin-0 resonance, $L_B=1$ and $L_R=0$. For $\pi^+\pi^-$ resonances with non-zero spin, $L_B$ can be 0, 1 or 2 (1, 2 or 3) for $L_R=1(2)$ and so on. We take the lowest $L_B$ as the default.  

The Blatt-Weisskopf barrier factors $F_B^{(L_B)}$ and  $F_R^{(L_R)}$ \cite{Blatt}  are 
 \begin{eqnarray}
F^{(0)} &=& 1, \nonumber \\
F^{(1)} &=& \frac{\sqrt{1+z_0}}{\sqrt{1+z}},\\
F^{(2)} &=& \frac{\sqrt{z_0^2+3z_0+9}}{\sqrt{z^2+3z+9}}. \nonumber
\end{eqnarray}
For the $B$ meson $z = r^2P_B^2$, where $r$, the hadron scale, is taken as 5.0 GeV$^{-1}$; for  the $R$ resonance $z = r^2P_R^2$, and $r$ is taken as 1.5 GeV$^{-1}$. In both cases $z_0= r^2P_0^2$ where $P_0$ is the decay daughter momentum at the pole mass, different for the \Bzb and  the resonance decay.

The angular term, $T_{\lambda}$, is obtained using the helicity formalism and is defined as 
\begin{equation}
 T_{\lambda} = d^J_{\lambda 0}(\theta_{\pi\pi}),
\end{equation}
where $d$ is the Wigner d-function \cite{PDG},
 $J$ is the resonance spin, $\theta_{\pi\pi}$ is the $\pi^+\pi^-$ resonance helicity angle which is defined as the angle of $\pi^+$ in the $\pi^+\pi^-$ rest frame with respect to the $\pi^+\pi^-$direction in the $\Bsb$ rest frame  and calculated from the other variables as
 \begin{equation}
\cos \theta_{\pi\pi} = \frac{\left[m^2(J/\psi \pi^+)-m^2(J/\psi \pi^-)\right]m(\pi^+\pi^-)}{4P_R P_B m_{B}}. \label{heli1}
\end{equation}
The $J/\psi$ helicity dependent term  $\Theta_{\lambda}(\theta_{J/\psi})$ is defined as
\begin{eqnarray}
 \Theta_{\lambda}(\theta_{J/\psi})&=& \sqrt{\sin^2\theta_{J/\psi}}\;\;\;\;\;\;\;\;\; \text{for}\;\; \text{helicity} = 0 \nonumber \\
 &=&\sqrt{\frac{1+\cos^2\theta_{J/\psi}}{2}}\;\; \text{for}\;\; \text{helicity} = \pm1. \label{heli2}
\end{eqnarray} 

The function $A_R(s_{23})$ describes the mass squared shape of the resonance $R$, that in most cases is a 
Breit-Wigner (BW) amplitude. Complications arise, however,  when a new decay channel opens close to the resonant mass. The proximity of a second threshold distorts the line shape of the amplitude.  This happens for the $f_0(980)$ because the $K^+K^-$ decay channel opens. Here we use a Flatt\'e model \cite{Flatte:1976xv}.  For non-resonant processes, the amplitude $A_R(s_{23})$ is constant over the variables $s_{12}$ and $s_{23}$, and has an angular dependence due to the $J/\psi$ decay.

 
The BW amplitude for a resonance decaying into two spin-0 particles, labeled as 2 and 3, is
\begin{equation}
A_R(s_{23})=\frac{1}{m^2_R-s_{23}-im_R\Gamma(s_{23})}~,
\end{equation}
where $m_R$ is the resonance mass, $\Gamma(s_{23})$ is its energy-dependent width that is parametrized as 
\begin{equation}
\Gamma(s_{23})=\Gamma_0\left(\frac{P_R}{P_{R_0}}\right)^{2L_R+1}\left(\frac{m_R}{\sqrt{s_{23}}}\right)F^2_R~.
\end{equation}
Here $\Gamma_0$ is the decay width when the invariant mass of the daughter combinations is equal to $m_R$.

 
The Flatt\'e model is parametrized as 
\begin{equation}
A_R(s_{23})=\frac{1}{m_R^2-s_{23}-im_R(g_{\pi\pi}\rho_{\pi\pi}+g_{KK}\rho_{KK})}.
\end{equation}
The constants  $g_{\pi\pi}$ and $g_{KK}$ are the $f_0(980)$ couplings to $\pi^+\pi^-$ and $K^+K^-$ final states respectively.
 The $\rho$ factors are given by Lorentz-invariant phase space
\begin{eqnarray}
\rho_{\pi\pi} &=& \frac{2}{3}\sqrt{1-\frac{4m^2_{\pi^{\pm}}}{m^2(\pi^+\pi^-)}}+\frac{1}{3}\sqrt{1-\frac{4m^2_{\pi^{0}}}{m^2(\pi^+\pi^-)}}\label{flatte1}, \\ 
\rho_{KK} &=& \frac{1}{2}\sqrt{1-\frac{4m^2_{K^{\pm}}}{m^2(\pi^+\pi^-)}}+\frac{1}{2}\sqrt{1-\frac{4m^2_{K^{0}}}{m^2(\pi^+\pi^-)}}.\label{flatte2}
\end{eqnarray}


The non-resonant amplitude is parametrized as
\begin{equation}
\mathcal{A}(s_{12},s_{23}, \theta_{J/\psi}) = \frac{P_B}{m_B} \sqrt{\sin^2\theta_{J/\psi}}.
\end{equation}

\subsection{Detection efficiency}
\label{sec:mc}

The detection efficiency is determined from a sample of one million $\Bsb\rightarrow J/\psi\pi^+\pi^-$ Monte Carlo  (MC) events  that are generated flat in phase space with $J/\psi \rightarrow \mu^+\mu^-$, using \pythia \cite{Sjostrand:2006za} with a special LHCb parameter tune \cite{LHCb-PROC-2011-005}, and the LHCb detector simulation based on G{\sc eant}4 \cite{Agostinelli:2002hh} described in Ref~\cite{LHCb-PROC-2011-006}. After the final selections the MC has 78,470 signal events, reflecting an overall efficiency of $7.8\%$. The acceptance in $\cos\theta_{J/\psi}$ is uniform. 
 
Next we describe the acceptance in terms of the mass squared variables. Both  $s_{12}$ and $s_{13}$ range from  $10.2\;\; \rm GeV^2$  to $27.6\;\;\rm GeV^2$, where $s_{13}$ is defined below,  and thus are centered at 18.9\, GeV$^2$. 
We model the detection efficiency using the symmetric Dalitz plot observables
\begin{equation}
x= s_{12}-18.9~{\rm GeV}^2,~~~~{\rm and}~~~~  y=s_{13}-18.9~{\rm GeV}^2.
\end{equation}
These variables are related 
to $s_{23}$ as
\begin{equation}
s_{12}+s_{13}+s_{23}=m^2_B+m^2_{J/\psi}+m^2_{\pi^+}+m^2_{\pi^-}~.\label{conver}
\end{equation}

The detection efficiency is parametrized as a symmetric 4th order polynomial function given by
\begin{eqnarray}
\varepsilon(s_{12}, s_{23})&=& 1+\varepsilon_1(x+y)+\varepsilon_2(x+y)^2+\varepsilon_3xy+\varepsilon_4(x+y)^3
+\varepsilon_5 xy(x+y)\nonumber \\
&&+\varepsilon_6(x+y)^4+\varepsilon_7 xy(x+y)^2+\varepsilon_8 x^2y^2,
\end{eqnarray}
where the $\varepsilon_i$ are the fit parameters. 

\begin{figure}[t]
\begin{center}
     \includegraphics[scale=0.55]{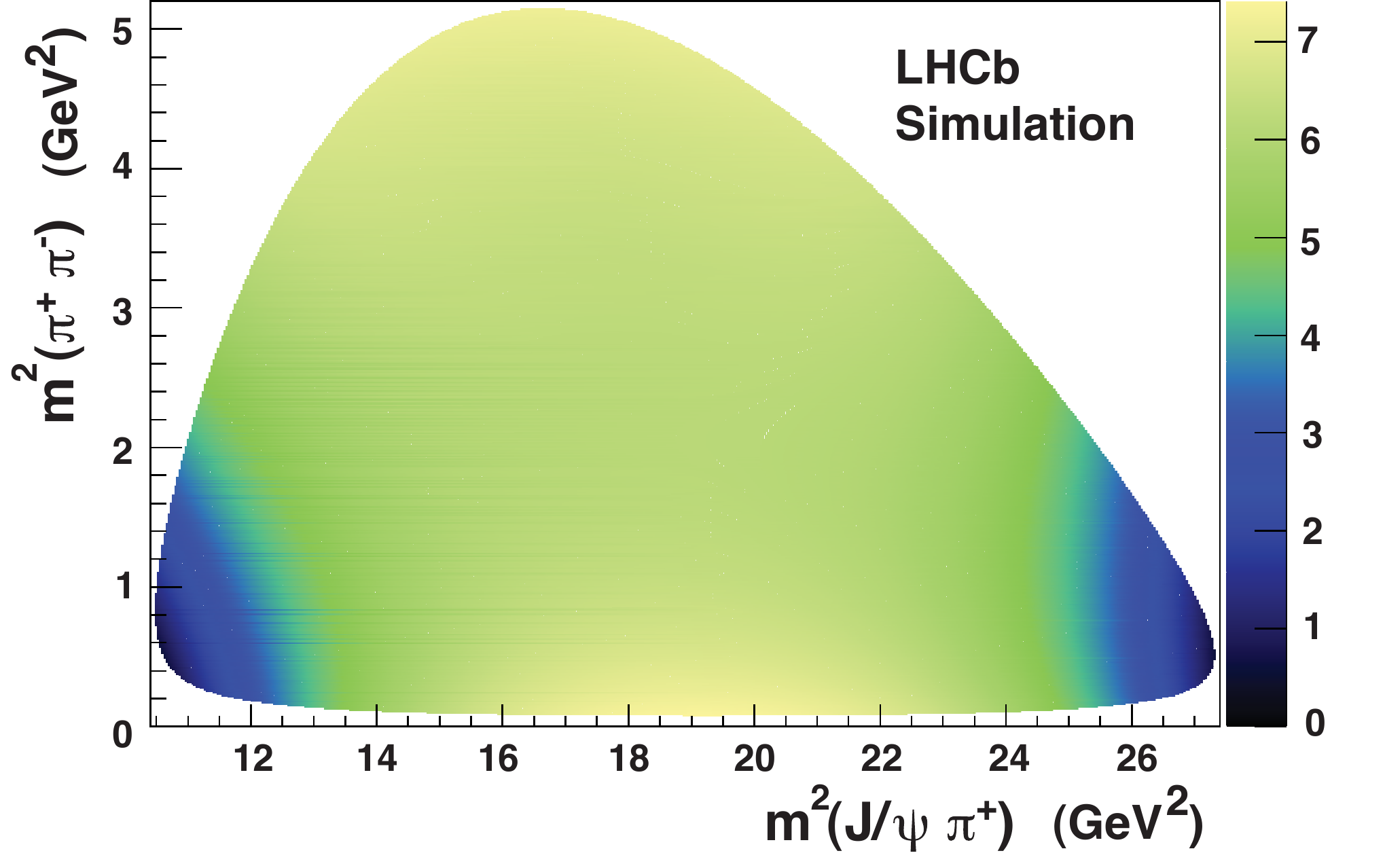}
\end{center}\label{eff1}
\vskip -0.5cm
\caption{Parametrized detection efficiency as a function
  of  $s_{23}\equiv m^2(\pi^+\pi^-)$ versus $s_{12}\equiv m^2(J/\psi\pi^+)$. The scale is arbitrary.}
\end{figure}

\begin{figure}[htb]
\begin{center}
    \includegraphics[width=6 in]{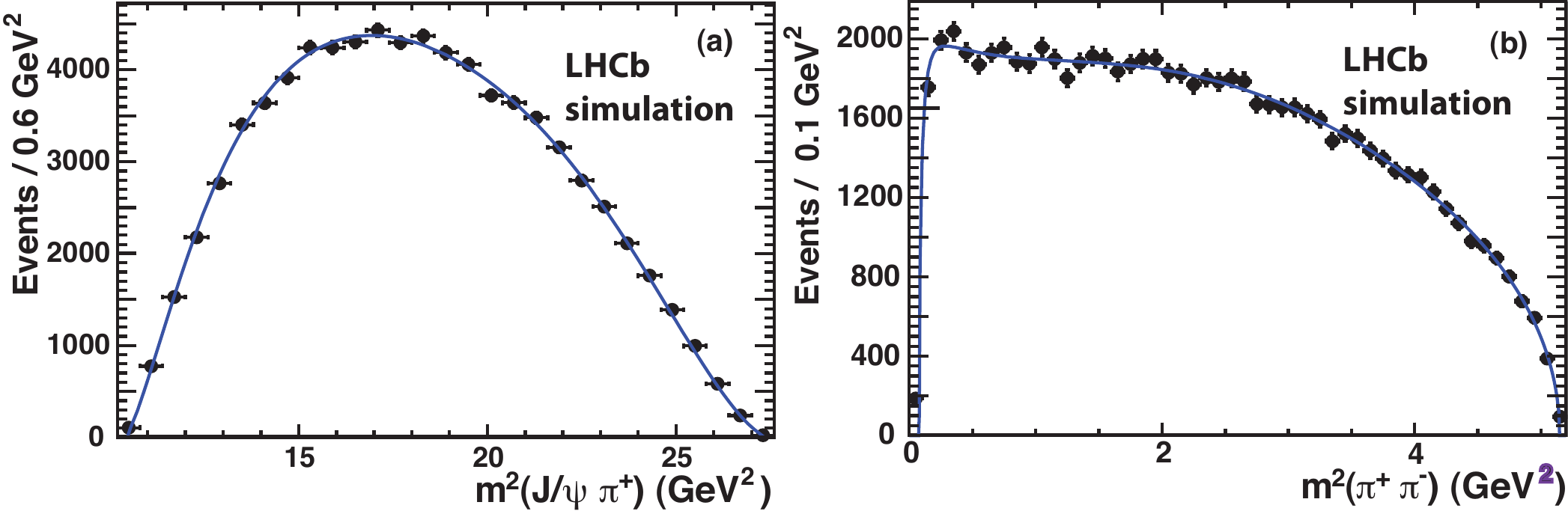}
\end{center}\label{eff2}
\vskip -0.5cm
\caption{Projections of invariant mass squared of (a) $s_{12}\equiv m^2(J/\psi \pi^+)$ and (b) $s_{23}\equiv m^2(\pi^+\pi^-)$ of the MC Dalitz plot used to measure the efficiency parameters. The points represent the MC generated event distributions and the curves the polynomial fit.}
\end{figure}

The fitted polynomial function is shown in Fig.~\ref{eff1}.  
The projections of the fit used to measure the efficiency parameters are shown in Fig.~\ref{eff2}.  The  efficiency shapes are well described by the parametrization. 

\begin{figure}[b]
\begin{center}
    \includegraphics[width=6 in]{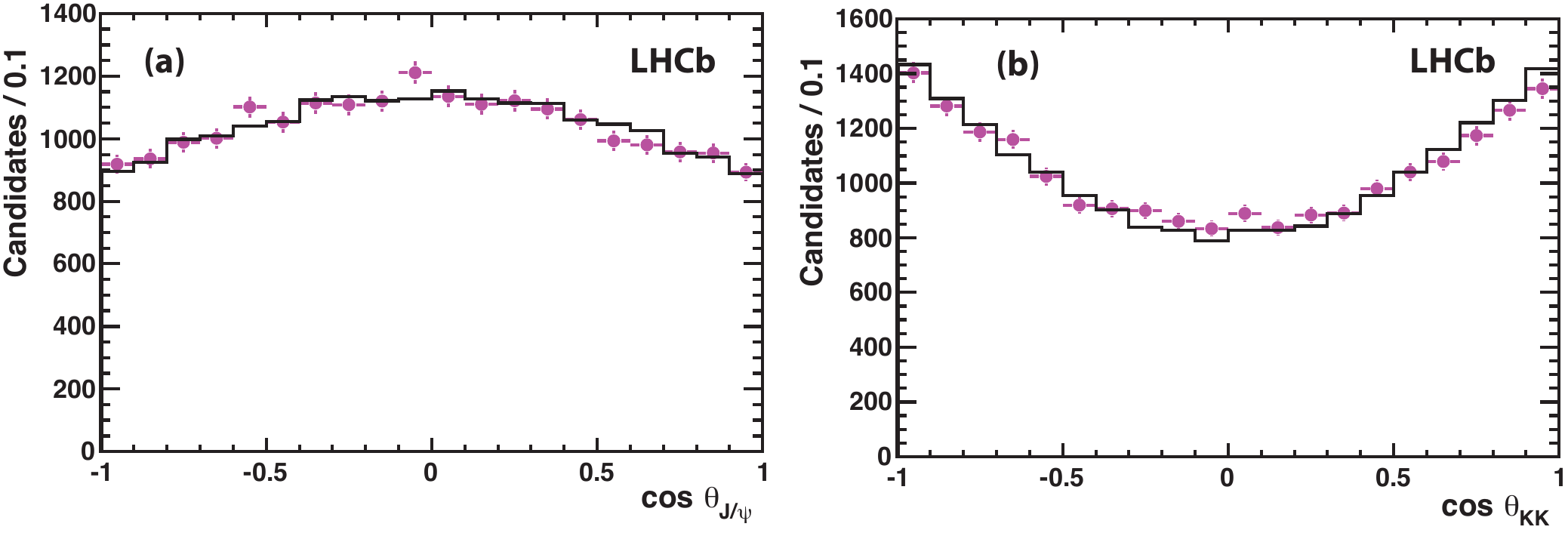}
\end{center}\label{cos_theta_Phi}
\vspace{-6mm}
\caption{ Distributions of  (a)
 $\cos\theta_{J/\psi}$, (b) $\cos\theta_{KK}$ for $J/\psi\phi$ background subtracted data (points) compared with the MC simulation (histogram). }
\end{figure}

To check the detection efficiency we compare our simulated $J/\psi\phi$ events with our measured $J/\psi\phi$ helicity distributions. The events are generated in the same manner as for $J/\psi\pi^+\pi^-$. Here we use the measured helicity amplitudes of $\left|A_{||}(0)\right|^2=0.231$ and $\left|A_{0}(0)\right|^2=0.524$ \cite{CDF:2011af}.  The background subtracted $J/\psi\phi$ angular distributions,  $\cos\theta_{J/\psi}$ and $\cos\theta_{KK}$, defined in the same manner as for the $J/\psi\pi^+\pi^-$ decay, are compared in Fig.~\ref{cos_theta_Phi} with the MC simulation.  The $\chi^2$/ndf =389/400 is determined by binning the angular distributions in two dimensions. The p-value is 64.1\%.
The excellent agreement gives us confidence that the simulation accurately predicts the acceptance.

\subsection{Background composition}

\begin{figure}[htb]
\begin{center}
    \includegraphics[width=6 in]{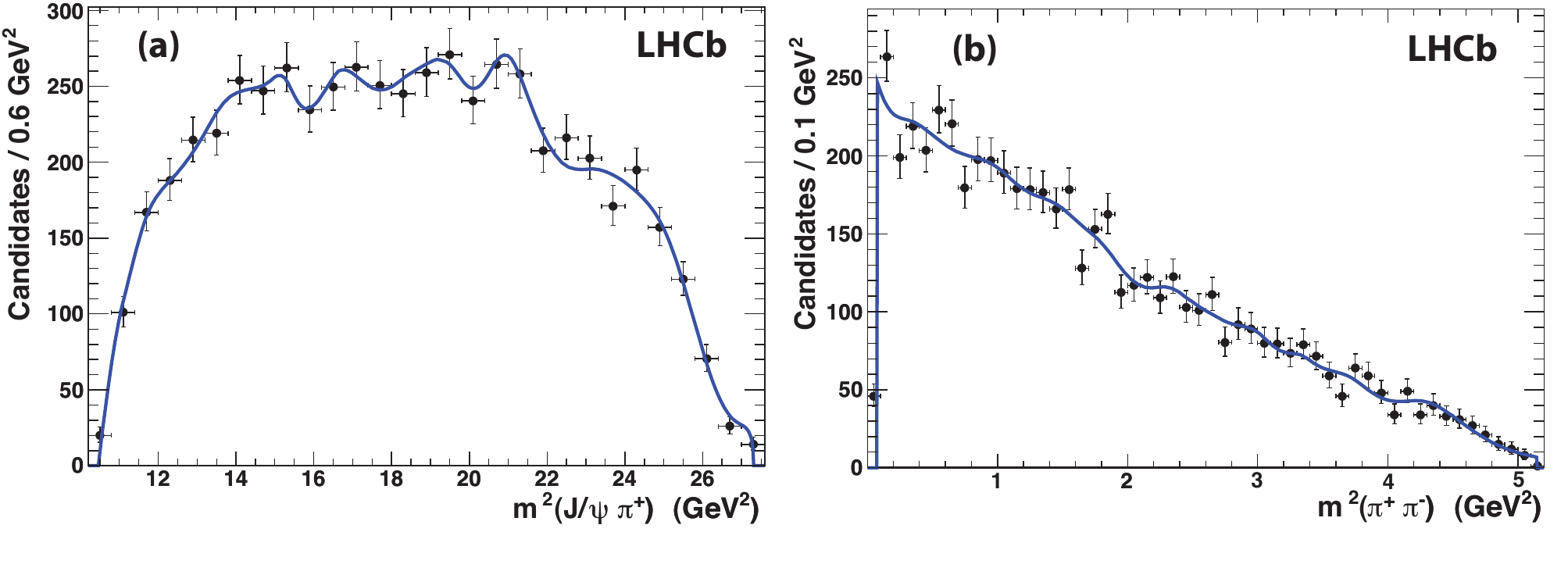}
\end{center}\label{bkg2}
\vskip -0.5cm
\caption{Projections of invariant mass squared of (a) $s_{12}\equiv m^2(J/\psi \pi^+)$ and (b) $s_{23}\equiv m^2(\pi^{\pm}\pi^{\pm})$ of the background Dalitz  plot. }
\end{figure}

\begin{figure}[htb]
\begin{center}
\includegraphics[scale=0.5]{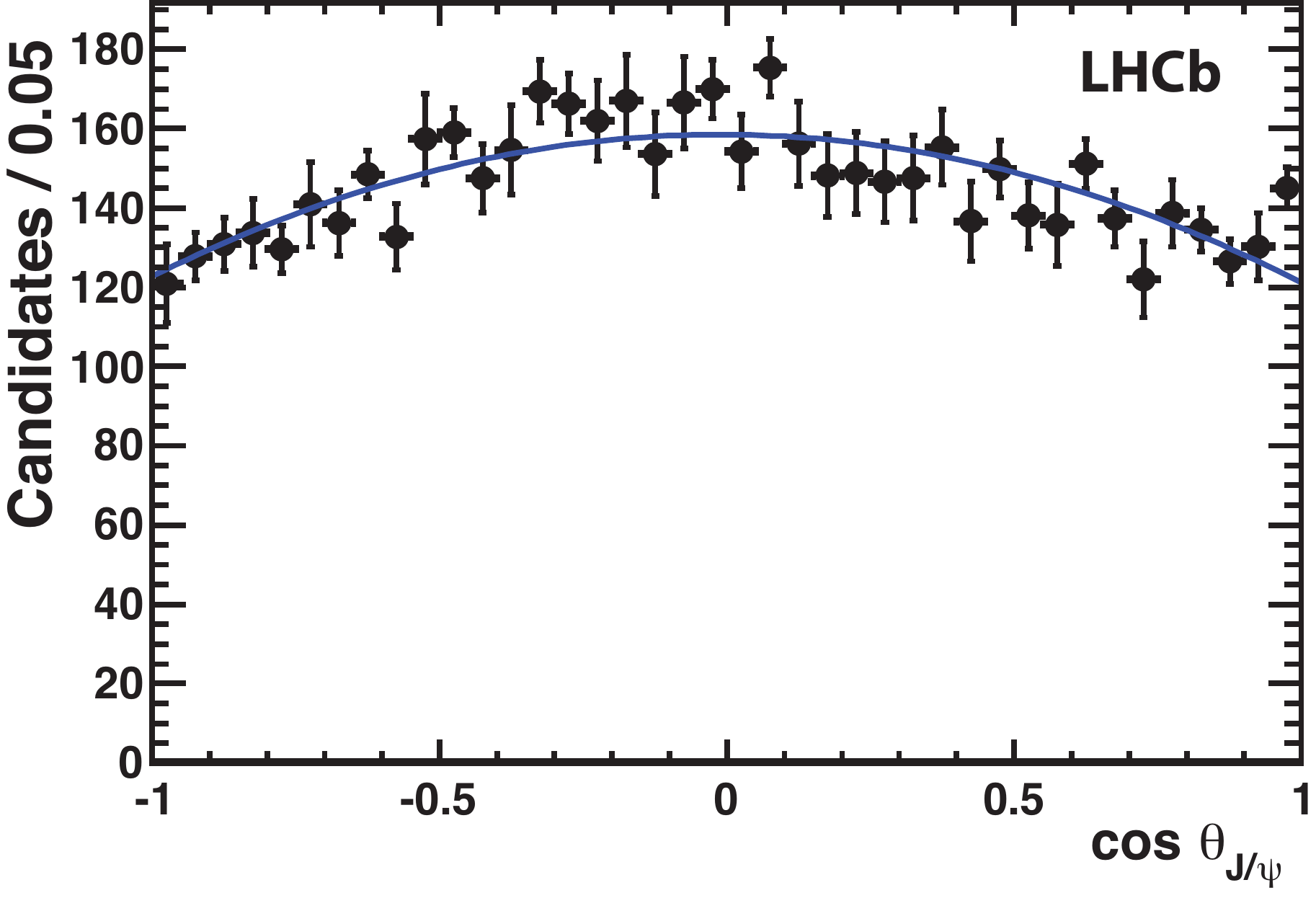}
\end{center}\label{bkg3}
\vskip -1cm
\caption{The $\cos\theta_{J/\psi}$ distribution of the background and the fitted function $1+\alpha\cos\theta_{J/\psi}+\beta \cos^{2}\theta_{J/\psi}$.}
\end{figure}

The main background source is taken from the wrong-sign combinations
 within $\pm20$\,MeV of the $\Bsb$ mass peak. In addition, an extra 4.5\% contribution from
 combinatorial background formed by \jpsi  and random
 $\rho(770)$, which cannot be present in wrong-sign combinations, is included using a MC sample.  The level is determined by measuring the background yield as a function of $\pi^+\pi^-$ mass. The background model is parametrized as
\begin{equation}
B(s_{12}, s_{23}, \theta_{J/\psi})=B_1(s_{12},s_{23})\times \left(1+\alpha\cos\theta_{J/\psi}+\beta \cos^{2}\theta_{J/\psi}\right),
\end{equation}
where the first part $B_1(s_{12},s_{23})$ is modeled using the technique of multiquadric radial basis functions \cite{Allison:1993dn}. These functions provide a useful way to parametrize
 multi-dimensional data giving sensible non-erratic behaviour and  yet they follow significant variations in a smooth and faithful way. They are useful in this analysis in providing a modeling of the decay angular distributions in the resonance regions.
Figure~\ref{bkg2} shows the mass squared projections from the fit. The $\chi^2/{\rm ndf}$ of the fit is 182/145. We also used such functions with half the number of parameters and the changes were insignificant. The second part $\left(1+\alpha\cos\theta_{J/\psi}+\beta \cos^{2}\theta_{J/\psi}\right)$ is a function of $J/\psi$ helicity angle. The $\cos\theta_{J/\psi}$ distribution of background is shown in Fig.~\ref{bkg3}, fit with the function $1+\alpha\cos\theta_{J/\psi}+\beta \cos^{2}\theta_{J/\psi}$ that determines the parameters $\alpha= -0.0050\pm0.0201$  and $\beta=-0.2308\pm0.0036$.


\section{Final state composition}\label{fit}
\label{sec:Results1}

\subsection{Resonance models}
 
To study the resonant structures of the decay $\Bsb\rightarrow J/\psi \pi^+\pi^-$ we use 13,424 candidates with invariant mass within $\pm20$ MeV of the $\Bsb$ mass peak. This includes both signal and background. 
 Possible resonance candidates in the decay $\Bsb\rightarrow J/\psi \pi^+\pi^-$ are listed in Table \ref{reso1}. 
\begin{table}[h!]
\begin{center}
\caption{Possible resonance candidates in the $\Bsb\rightarrow J/\psi \pi^+\pi^-$ decay mode.}
\begin{tabular}{cccc}
\hline
Resonance & Spin & Helicity & Resonance \\
&&& formalism \\
\hline
$f_0(600)$  & 0 & 0 & BW \\
$\rho(770)$ & 1 & $0,\pm 1$ & BW \\

$f_0(980)$ & 0 & 0 & Flatt\'e \\

$f_2(1270)$ & 2 &  $0,\pm 1$ & BW \\

$f_0(1370)$ & 0 & 0 & BW \\

$f_0(1500)$ & 0 & 0 & BW \\
\hline
\end{tabular}\label{reso1}
\end{center}
\end{table}
To understand  what resonances are likely to contribute, it is important to realize that the $s\bar{s}$ system in Fig.~\ref{feyn1} is isoscalar ($I=0$) so when it produces a single meson it must have zero isospin, resulting in a symmetric isospin wavefunction for the two-pion system. Since the two-pions must be in an overall symmetric state, they must have even total angular momentum. 
In fact we only need to consider spin-0 and spin-2 particles as there are no known spin-4 particles in the kinematically accessible mass range below 1600 MeV. The particles that could appear are spin-0 $f_0(600)$, spin-0 $f_0(980)$, spin-2 $f_2(1270)$, spin-0 $f_0(1370)$ and spin-0 $f_0(1500)$. 
Diagrams of higher order than the one shown in Fig.~\ref{feyn1} could result in the production of isospin-one $\pi^+\pi^-$ resonances, thus we use the $\rho(770)$ as a test of the presence of these higher order processes.
  
We proceed by fitting with a single $f_0(980)$, established from earlier measurements \cite{Aaij:2011fx}, and adding single resonant components until acceptable fits are found. Subsequently, we try the addition of other resonances. The models used are listed in Table~\ref{tab:modelsused}.
\begin{table}[h!]
\begin{center}
\caption{Models used in data fit.}
\begin{tabular}{ll}
\hline
Name & Components\\\hline
{Single R} & $f_0(980)$\\
{2R}&$f_0(980)+f_0(1370)$\\
{3R}&$f_0(980)+f_0(1370)+f_2(1270)$\\
{3R+NR}&$f_0(980)+f_0(1370)+f_2(1270) + ~$non-resonant \\
{3R+NR +} {$\rho(770)$}& $f_0(980)+f_0(1370)+f_2(1270) + ~$non-resonant $+\rho(770)$ \\
{3R+NR +}{$f_0(1500)$}&$f_0(980)+f_0(1370)+f_2(1270) + ~$non-resonant $+f_0(1500)$ \\
{3R+NR +}{$f_0(600)$}& $f_0(980)+f_0(1370)+f_2(1270) + ~$non-resonant $+f_0(600)$\\
\hline
\end{tabular}\label{tab:modelsused}
\end{center}
\end{table}

The masses and widths of the BW resonances are listed in Table~\ref{PDG_param}. When used in the fit they are fixed to these values, except for the $f_0(1370)$, for which they are not well measured, and thus are allowed to vary using their quoted errors as constraints in the fits, taking the errors as being Gaussian.

Besides the mass and width, the Flatt\'e resonance shape has two additional parameters $g_{\pi\pi}$ and $g_{KK}$, which are also allowed to vary in the fit. Parameters of the non-resonant amplitude are also allowed to vary.
One magnitude and one phase in each helicity grouping have to be fixed, since the overall normalization is related to the signal yield, and only relative phases are physically meaningful. The normalization and phase of $f_0(980)$ are fixed to 1 and 0 respectively. The phase of $f_2(1270)$, with helicity $=\pm1$ is also fixed to zero when it is included. All background and efficiency parameters are held static in the fit. 
\begin{table}[h]
\label{tab:resparam}
\begin{center}
\caption{Breit-Wigner resonance parameters.}\label{PDG_param}
\begin{tabular}{cccc}
\hline
 Resonance &Mass (MeV) & Width (MeV) & Source \\
 \hline
 $f_0(600)$~~\! & $513\pm32$ & $335\pm67$ & CLEO \cite{Muramatsu:2002jp}\\
$\rho(770)$~~ & $775.5\pm0.3$ &$149.1\pm0.8$&PDG \cite{PDG}\\ 
$f_2(1270)$ & $1275\pm1$ & $185\pm3$&PDG \cite{PDG} \\
 $f_0(1370)$ &  $1434\pm20$& $172\pm33$ &E791 \cite{Aitala:2000xt}\\
$f_0(1500)$ & $1505\pm6$ & $109\pm$7&PDG \cite{PDG}\\
\hline
\end{tabular}
\end{center}
\end{table}
 
To determine the complex amplitudes in a specific model, the data are fitted maximizing the unbinned likelihood given as
\begin{equation}
{\cal{L}}=\prod_{i=1}^NF\left(s_{12}^i,s_{23}^i,\theta_{\jpsi}^i\right),
\end{equation}
where $N$ is the total number of events, and $F$ is the total PDF defined
in Eq.~\ref{eq:F}.
The PDF is constructed from the signal fraction $f_{\rm sig}$, efficiency model $\varepsilon (s_{12}, s_{23})$, background model $B(s_{12}, s_{23},\theta_{J/\psi})$ and the signal model $S(s_{12}, s_{23},\theta_{J/\psi})$.
The PDF needs to be normalized. This is accomplished by first normalizing the $J/\psi$ helicity dependent part by analytical integration, and then for the mass dependent part using numerical integration over 500$\times$500 bins.
 
 \subsection{Fit results}

\begin{figure}[!t]
  \begin{center}
     \includegraphics[width=6.2in]{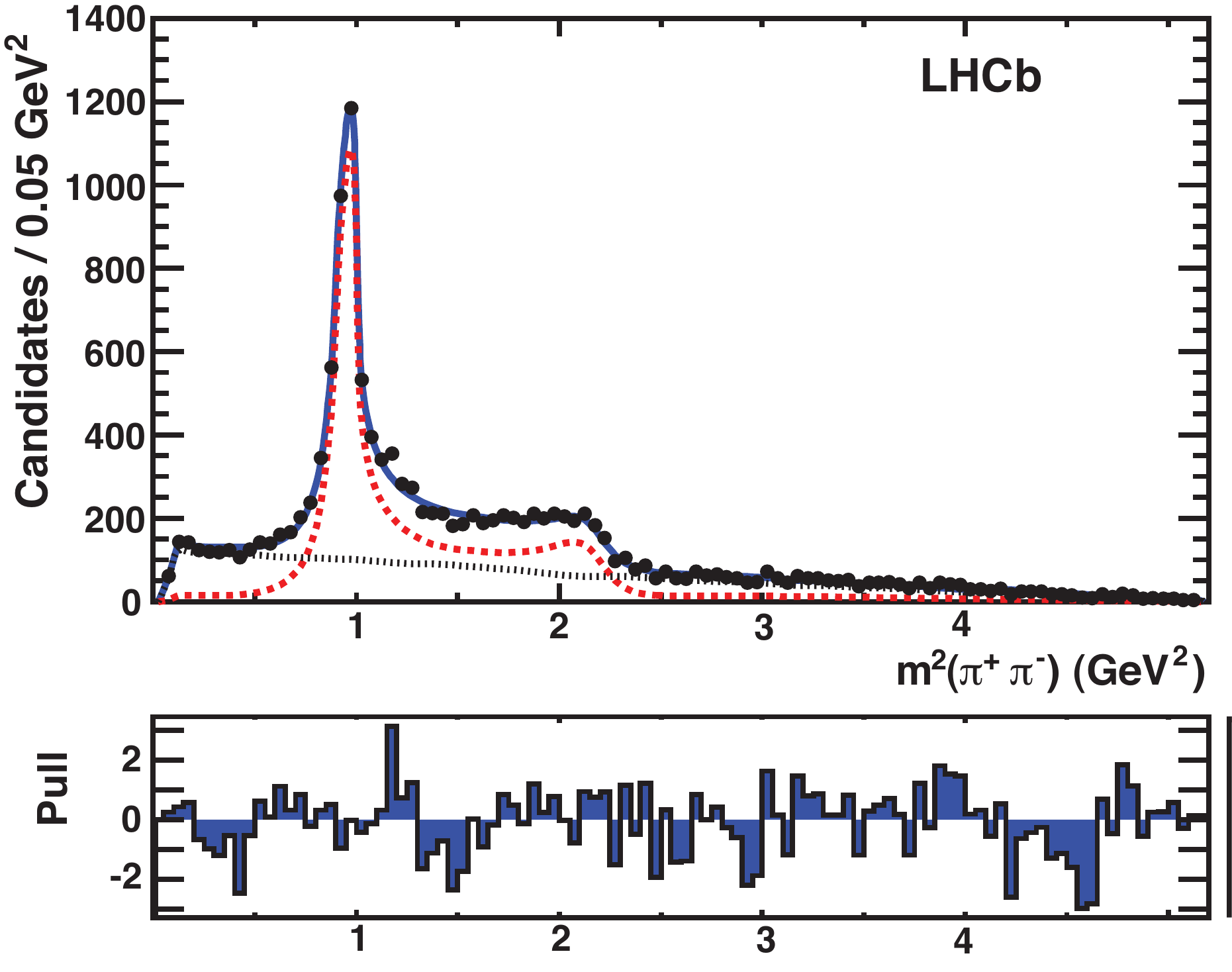}
 \caption{Dalitz fit projections of $m^2(\pi^+ \pi^-)$ fit with 3R+NR for the preferred model. The points with error bars are data, the signal fit is shown with a (red) dashed line, the background with a (black) dotted line, and the (blue) solid line represents the total. The normalized residuals in each bin are shown below, defined as the difference between the data and the fit divided by the error on the data. }  \label{RM4}
  \end{center}
\end{figure}

\begin{figure}[!t]
   \begin{center}
      \includegraphics[width=6.2in]{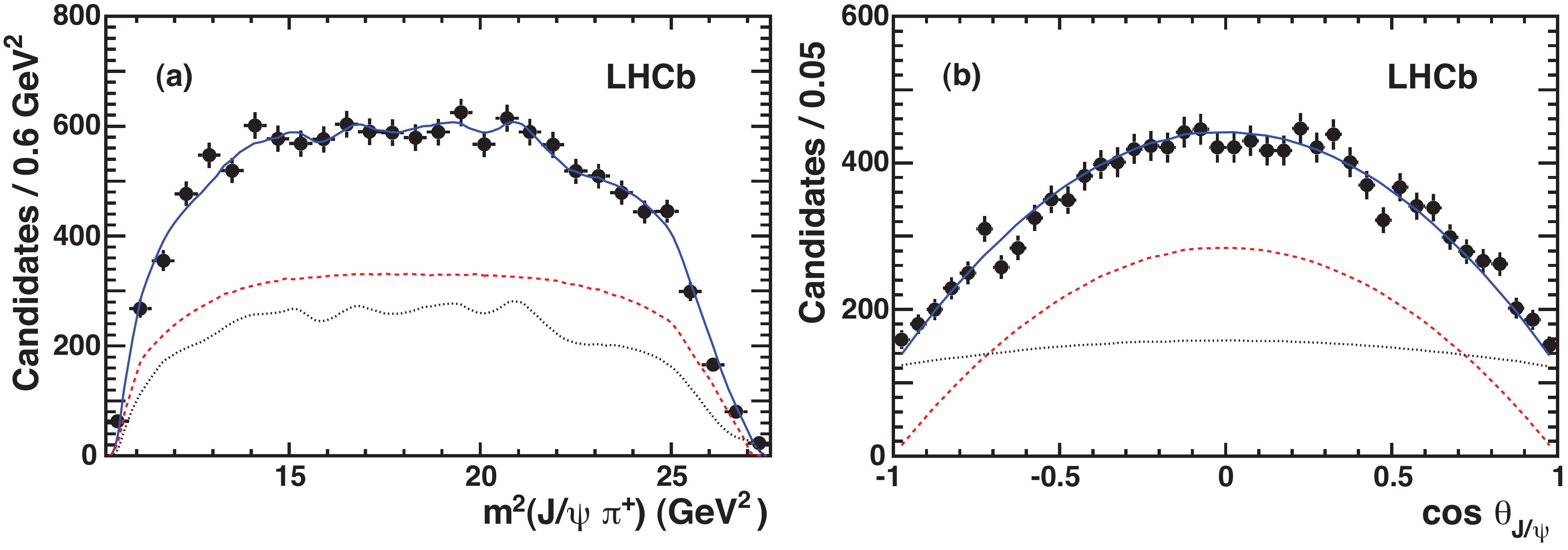}
       \caption{Dalitz fit projections of (a) $s_{12}\equiv m^2(J/\psi \pi^{+})$ and (b) $\cos \theta_{J/\psi}$ fit with the 3R+NR  preferred model. The points with error bars are data, the signal fit is shown with a (red) dashed line, the background with a (black) dotted line, and the (blue) solid line represents the total.}
       \label{8R1-R3}
  \end{center}
\end{figure}

In order to compare the different models quantitatively an estimate of the goodness of fit is calculated from 3D partitions of the one angular and two mass-squared variables. We use the Poisson likelihood $\chi^2$ \cite{Baker:1983tu} defined as 
\begin{equation}
\chi^2=2\sum_{i=1}^{N_{\rm bin}}\left[  x_i-n_i+n_i \text{ln}\left(\frac{n_i}{x_i}\right)\right],
\end{equation}
where $n_i$ is the number of events in the three dimensional bin $i$ and $x_i$ is the expected number of events in that bin according to the fitted likelihood function. A total of $N_{\rm bin}=1356$ bins are used to calculate the $\chi^2$, using the variables $m^2(J/\psi\pi^+)$,  $m^2(\pi^+\pi^-)$, and $\cos\theta_{J/\psi}$.  The $\chi^2/\text{ndf}$ and the negative of the logarithm of the likelihood, $\rm -ln\mathcal{L}$, of the fits are given in Table~\ref{RMchi2}. There are two solutions of almost equal likelihood for the 3R+NR model. Based on a detailed study of angular distributions (see Section~\ref{sec:hel-dist}) we choose one of these solutions and label it as ``preferred". The other solution is called ``alternate." We will use the differences between these to assign systematic uncertainties to the resonance fractions. 
\begin{table}[h!t!p!]
\begin{center}
\caption{$\chi^2/\text{ndf}$ and $\rm -ln\mathcal{L}$ of different resonance models.}
\begin{tabular}{lccc}
\hline
Resonance model & $\rm -ln\mathcal{L}$& $\chi^2/\text{ndf}$ & Probability (\%) \\
\hline
Single R &59269& 1956/1352 & 0\\
2R &59001 &1498/1348 & 0.25\\
3R &58973 &1455/1345 & 1.88 \\
3R+NR (preferred) &58945 & 1415/1343 & 8.41\\
3R+NR (alternate) &58946 & 1414/1343 & 8.70\\
\hline
3R+NR + $\rho(770)$ (preferred) &58945 &1418/1341&7.05\\
3R+NR + $\rho(770)$ (alternate) &58944 &1416/1341&7.57\\
3R+NR + $f_0(1500)$ (preferred)   &58943 & 1416/1341&7.57\\
3R+NR + $f_0(1500)$ (alternate)  &58941 & 1407/1341&\!10.26~~\\
3R+NR + $f_0(600)$ (preferred)   &58935 & 1409/1341 &9.60\\
3R+NR + $f_0(600)$ (alternate)  &58937 & 1412/1341&8.69\\
\hline
\end{tabular}
\label{RMchi2}
\end{center}
\end{table}
The probability is improved noticeably adding components up to 3R+NR. Figure~\ref{RM4} shows the preferred model projections of $m^2(\pi^+\pi^-)$ for 
the preferred model including only the 3R+NR components. The projections for the other considered models are indiscernible.
The preferred model projections of  $m^2(J/\psi \pi^{+})$ and $\cos \theta_{J/\psi}$ are shown in Fig.~\ref{8R1-R3} for the preferred model 3R+NR fit. The projections of the other preferred model fits including the additional resonances are almost identical.



While a complete description of the decay is given in terms of the fitted amplitudes and phases, knowledge of the contribution of each component can be
summarized by defining a fit fraction, ${\cal{F}}^R_{\lambda}$. To determine ${\cal{F}}^R_{\lambda}$ we integrate the squared amplitude of $R$ over the Dalitz plot. The yield is then normalized by integrating the entire signal function over the same area. Specifically,  
\begin{equation}
{\cal{F}}^R_{\lambda}=\frac{\int\left| a^R_{\lambda} e^{i\phi^R_{\lambda}} \mathcal{A}_{\lambda}^{R}(s_{12},s_{23},\theta_{J/\psi})\right|^2 ds_{12}\;ds_{23}\;d\cos\theta_{J/\psi}}{\int S(s_{12},s_{23},\theta_{J/\psi})  ~ds_{12}\;ds_{23}\;d\cos\theta_{J/\psi}}.
\end{equation}
Note that the sum of the fit fractions is not necessarily unity due to the potential presence of interference between two resonances. Interference term fractions are given by
\begin{equation}
\label{eq:inter}
{\cal{F}}^{RR'}_{\lambda}=2\mathcal{R}e\left(\frac{\int a^R_{\lambda}\; a^{R'} _{\lambda}e^{i(\phi^R_{\lambda}-\phi^{R'}_{\lambda})} \mathcal{A}_{\lambda}^{R}(s_{12},s_{23},\theta_{J/\psi}) {\mathcal{A}_{\lambda}^{R'}}^{*}(s_{12},s_{23},\theta_{J/\psi}) ds_{12}\;ds_{23}\;d\cos\theta_{J/\psi}}{\int S (s_{12},s_{23},\theta_{J/\psi}) ~ds_{12}\;ds_{23}\;d\cos\theta_{J/\psi}}\right),
\end{equation}
and
\begin{equation}
\sum_{\lambda}\left(\sum_R {\cal{F}}^R_{\lambda}+\sum_{RR'} {\cal{F}}^{RR'}_{\lambda}\right) =1.
\end{equation}
If the Dalitz plot has more destructive interference than constructive interference, the total fit fraction will be greater than one.  Note that, interference  between different spin-$J$ states vanishes because the $d^J_{\lambda0}$ angular functions in $\mathcal{A}^R_{\lambda}$ are orthogonal.
 
The determination of the statistical errors of the fit fractions is difficult because they depend on the statistical errors of every fitted magnitude and phase. A toy Monte Carlo approach is used. We perform 500 toy  experiments: each sample is generated according to the model PDF, input parameters are taken from the fit to the data.  The correlations of fitted parameters are also taken into account. For each toy experiment the fit fractions are calculated. 
The distributions of the obtained fit fractions are described by Gaussian functions. The r.m.s. widths of the Gaussians are taken as the statistical errors on the corresponding parameters. The fit fractions are listed in Table \ref{ff3}.

\begin{table}[h!t!p!]
\begin{center}
\caption{Fit fractions (\%) of contributing components for the preferred model. For P- and D-waves $\lambda$ represents the final state helicity. Here $\rho$ refers to the $\rho(770)$ meson.}
\begin{tabular}{lcccc}
\hline
~~Components & 3R+NR & 3R+NR+$\rho$& 3R+NR+$f_0(1500)$ & 3R+NR+$f_0(600)$\\
\hline
$f_0(980)$ &$107.1\pm3.5$~~\!&$104.8\pm3.9$&$73.0\pm5.8$& $115.2\pm5.3$\\
$f_0(1370)$ &$32.6\pm4.1$&~~\!$32.3\pm3.7$&$114\pm14$&~$34.5\pm4.0$\\
$f_0(1500)$ &-&-&$15.0\pm5.1$&-\\
$f_0(600)$ &-&-&-&~~$4.7\pm2.5$ \\
NR &$12.84\pm2.32$&~~\!$12.2\pm2.2$&$10.7\pm2.1$&~\!$23.7\pm3.6$ \\
$f_2(1270)$, $\lambda=0$ &~$0.76\pm0.25$&~~~$0.77\pm0.25$&~$1.07\pm0.37$&~~$0.90\pm0.31$\\
$f_2(1270)$, $|\lambda|=1$ &~$0.33\pm1.00$&~~~$0.26\pm1.12$&~$1.02\pm0.83$&~~$0.61\pm0.87$\\         
$\rho$, $\lambda=0$ & -&~~~$0.66\pm0.53$&-&-\\
$\rho$, $|\lambda|=1$ &-&~~~$0.11\pm0.78$&-&-\\\hline
Sum&  
$153.6\pm6.0$~&$151.1\pm6.0$&$214.4\pm15.7$&$179.6\pm8.0$\\
\hline
   $\rm -ln\mathcal{L}$&58945 &58944&58943&58935 \\
   $\chi^2$/ndf&1415/1343 &1418/1341&1416/1341&1409/1341\\
{\small Probability(\%)}& 8.41 &7.05&7.57&9.61\\
\hline
\end{tabular}\label{ff3}
\end{center}
\end{table}

The 3R+NR fit describes the data well. For models adding more resonances, the additional components never have more than 3 standard deviation ($\sigma$) significance, and the fit likelihoods are only slightly improved.
In the 3R+NR solution all the components have more than $3\sigma$ significance, except for the $f_2(1270)$ where we allow the helicity $\pm$1 components since the helicity 0 component is significant. In all cases, we find the dominant contribution is S-wave which agrees with our previous less sophisticated analysis \cite{LHCb:2011ab}. The D-wave contribution is small. The P-wave contribution is consistent with zero, as expected.
The fit fractions from the alternate model are listed in Table~\ref{ff1}. There are only small changes in the $f_2(1270)$  and  $\rho(770)$ components. 
\begin{table}[h!]
\begin{center}
\caption{Fit fractions (\%) of contributing components from different models for the alternate solution. For P- and D-waves $\lambda$ represents the final state helicity. Here $\rho$ refers to the $\rho(770)$ meson.}
\begin{tabular}{lcccc}
\hline
~~Components & 3R+NR & 3R+NR+$\rho$& 3R+NR+$f_0(1500)$ & 3R+NR+$f_0(600)$\\
\hline
$f_0(980)$ &\!$100.8\pm2.9$ &$99.2\pm4.2$&$96.9\pm3.8$&$111\pm15$ \\
$f_0(1370)$ &~~$7.0\pm0.9$&~~\!$6.9\pm0.9$&~~\!$3.0\pm1.7$&~~\!$8.0\pm1.1$\\
$f_0(1500)$ &- &-&~~\!$4.7\pm1.7$&-\\
$f_0(600)$ & -&-&-&~~\!$4.3\pm2.3$\\
NR & ~$13.8\pm2.3$ &~\!$13.4\pm2.7$&~\!\!$13.4\pm2.4$&~\!\!$24.7\pm3.9$\\
$f_2(1270)$, $\lambda=0$ &~~~\!$0.51\pm0.14$ &~~$0.52\pm0.14$&~~\!$0.50\pm0.14$&~~\!$0.51\pm0.14$\\
$f_2(1270)$, $|\lambda|=1$ &~~~\!$0.24\pm1.11$ &~~$0.19\pm1.38$&~~\!$0.63\pm0.84$&~~\!$0.48\pm0.89$\\         
$\rho$, $\lambda=0$ & -&~~$0.43\pm0.55$&-&-\\
$\rho$, $|\lambda|=1$ &- &~~$0.14\pm0.78$&-&-\\\hline
Sum&  \!$122.4\pm4.0$&\!\!$120.8\pm5.3$ &\!$119.2\pm5.2$&~\! \!$148.7\pm15.5$\\
\hline
   $\rm -ln\mathcal{L}$&58946 &58945&58941&58937 \\
   $\chi^2$/ndf&1414/1343 &1416/1341&1407/1341&1412/1341\\
   {\small Probability(\%)}& 8.70 &7.57&10.26&8.69\\
\hline
\end{tabular}
\end{center}\label{ff1}
\end{table}


The fit fractions of the interference terms for the preferred and alternate models are computed using Eq.~\ref{eq:inter} and listed in Table~\ref{tab:inter}.
\begin{table}[h!t!p!]
\begin{center}
\caption{Fit fractions (\%) of interference terms for both solutions of the 3R+NR model.}
\begin{tabular}{lcc}
\hline
~~~Components& Preferred & Alternate \\
\hline
$f_0(980)$ + $f_0(1370)$& $-36.6\pm4.6$ & ~~$-5.4\pm2.3$ \\
$f_0(980)$ + NR & $-16.1\pm2.7$& $-23.6\pm2.6$ \\
$f_0(1370)$ + NR& ~~~$~0.8\pm1.0$  & ~~$~~6.6\pm0.8$\\\hline
Sum&$-53.6\pm5.5$& $-22.4\pm3.6$\\
\hline
\end{tabular}
\end{center}\label{tab:inter}
\end{table}


\subsection{Helicity distributions}
\label{sec:hel-dist}
Only S and D waves contribute to the  $\Bsb\rightarrow J/\psi \pi^+\pi^-$ final state in the $m(\pi^+\pi^-)$ region below 1550 MeV.  Helicity information is already included in the signal model via Eqs.~\ref{heli1} and \ref{heli2}.  
For a spin-0 $\pi^+\pi^-$ system
 $\cos\theta_{J/\psi}$ should be distributed as $1-\cos^2\theta_{J/\psi}$ and $\cos\theta_{\pi\pi}$ should be flat. To test our fits we examine the $\cos\theta_{J/\psi}$ and $\cos\theta_{\pi\pi}$ distribution in different regions of $\pi^+\pi^-$ mass.  The decay rate with respect to the cosine of the helicity angles is given by \cite{LHCb:2011ab}
\begin{eqnarray}
\frac{d\Gamma}{•d\cos\theta_{J/\psi}d\cos\theta_{\pi\pi}}=&&
\left|A_{00}+\frac{1}{2}A_{20}e^{i\phi}\sqrt{5}(3\cos^2\theta_{\pi\pi}-1)\right|^2\sin^2\theta_{J/\psi} \\ \nonumber
&& + \frac{1}{4}\left(\left|A_{21}\right|^2+\left|A_{2-1}\right|^2\right)\left(15\sin^2\theta_{\pi\pi}\cos^2\theta_{\pi\pi}\right)
\left(1+\cos^2\theta_{J/\psi}\right),
\end{eqnarray}
where $A_{00}$ is the S-wave amplitude, $A_{2i},\;i=-1,0,1$, the three D-wave amplitudes, and $\phi$ is the strong phase between $A_{00}$ and $A_{20}$ amplitudes. Non-flat distributions in $\cos\theta_{\pi\pi}$ would indicate interference between the S-wave and D-wave amplitudes.

\begin{figure}[htb]
\begin{center}
\includegraphics[width=6.2in]{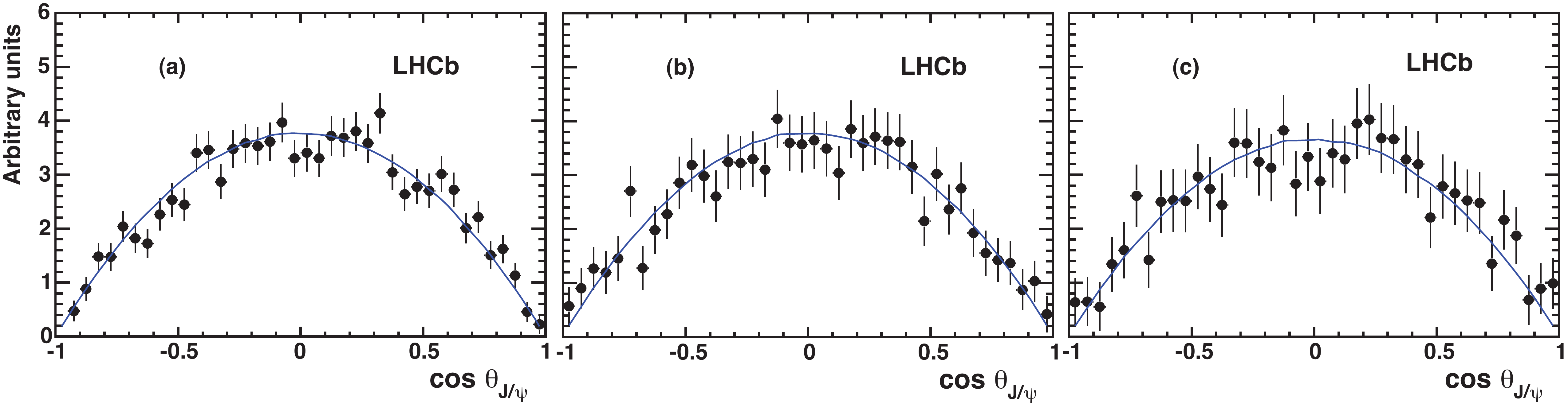}
\end{center}
\vspace{-6mm}
\caption{ Background subtracted and acceptance corrected $\cos \theta_{J/\psi}$ helicity distributions fit with the preferred model:  (a)
 in $f_0(980)$ mass region defined within $\pm90$ MeV of 980 MeV ($\chi^2$/ndf =39/40),  (b)
 in $f_2(1270)$ mass region defined within one full width of $f_2(1270)$ mass ($\chi^2$/ndf =25/40), (c) in $f_0(1370)$ mass region defined within one full width of $f_2(1370)$ mass ($\chi^2$/ndf = 24/40).  The points with error bars  are data and the solid blue lines show the fit from the 3R+NR model.}\label{helii1_3}
\end{figure}

\begin{figure}[htb]
\begin{center}
    \includegraphics[width=6.2in]{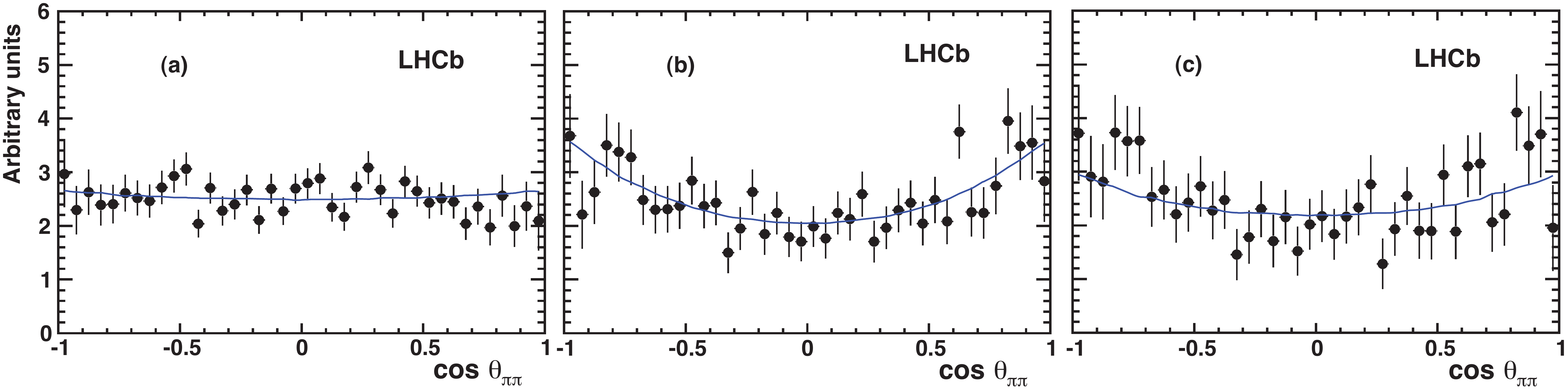}
\end{center}
\vspace{-6mm}
\caption{Background subtracted and acceptance corrected $\cos \theta_{\pi\pi}$ helicity distributions fit the preferred model:  (a)
 in $f_0(980)$ mass region defined within $\pm90$ MeV of 980 MeV ($\chi^2$/ndf =38/40),  (b)
 in $f_2(1270)$ mass region defined within one full width of $f_2(1270)$ mass ($\chi^2$/ndf = 32/40), (c) in $f_0(1370)$ mass region defined within one full width of $f_2(1370)$ mass ($\chi^2$/ndf =37/40).  The points with error bars  are data and the solid blue lines show the fit from the 3R+NR model.}\label{helii1_4}
\end{figure}

\begin{figure}[htb]
\begin{center}
    \includegraphics[width=5.in]{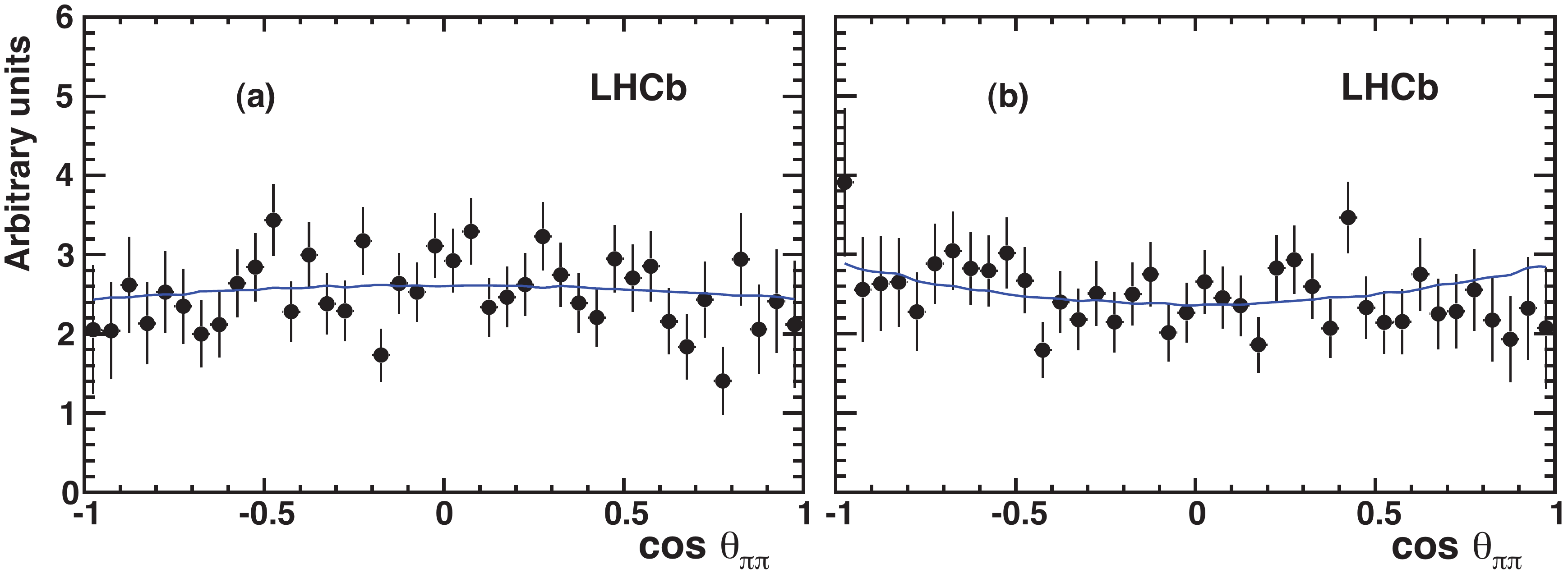}
\end{center}
\vspace {-6mm}
\caption{Background subtracted and acceptance corrected $\cos \theta_{\pi\pi}$ helicity distributions fit the preferred model:  (a)
 in $[-90,0]$ MeV of 980 MeV ($\chi^2$/ndf =41/40),  (b)
 in $[0,90]$ MeV of 980 MeV ($\chi^2$/ndf =31/40)}\label{helii1_5}
\end{figure}

To investigate the angular structure we then split the helicity distributions into three different $\pi^+\pi^-$ mass regions: one is the $f_0(980)$ region defined within $\pm90$ MeV of the $f_0(980)$ mass and the others are defined within one full width of the $f_2(1270)$ and $f_0(1370)$ masses, respectively (the width values are given in Table~\ref{tab:resparam}).
 The $\cos \theta_{J/\psi}$ and $\cos \theta_{\pi\pi}$ background-subtracted efficiency corrected distributions for these three different mass regions are presented in Figs.~\ref{helii1_3} and \ref{helii1_4}.  The distributions are in good agreement with the 3R+NR preferred signal model. 
Furthermore, splitting into two bins, $[-90,0]$ and $[0,90]$~MeV, we see different shapes,  because across the pole mass of $f_{0}(980)$, the $f_{0}(980)$'s phase changes by $\pi$. Hence the relative phase between $f_0(980)$ and the small D-wave in the two regions changes very sharply. This feature is reproduced well by the ``preferred" model and shown in Fig. \ref{helii1_5}. The ``alternate" model gives an acceptable, but poorer description.


\subsection{Resonance parameters}
 The fit results from the four-component best fit are listed in Table~\ref{fitres2} for both the preferred and alternate solutions.  The table summarizes the $f_0(980)$ mass, the Flatt\'e resonances parameters $g_{\pi\pi}$, $g_{KK}/g_{\pi\pi}$, $f_0(1370)$ mass and width and the phases of the contributing resonances.  

 \begin{table}[h!bt]
\begin{center}
\caption{Fit results from the 3R+NR model for both  the preferred and alternate solutions. $\phi$ indicates the phase with respect to the $f_0(980)$. For the $f_2(1270)$, $\lambda$ represents the final state helicity.}
\begin{tabular}{lcc}
\hline
The parameters & Preferred & Alternate\\
\hline
 $m_{f_0(980)}$(MeV) &$939.9\pm6.3$&$939.2\pm6.5$\\
 $g_{\pi\pi}$(MeV)   &~~\!$199\pm30$&~~\!$197\pm25$\\
 $g_{KK}/g_{\pi\pi}$ &~~~\!$3.0\pm0.3$&~~~\!$3.1\pm0.2$ \\
 $m_{f_0(1370)}$(MeV)&\!\!\!\!$1475.1\pm6.3$&\!\!\!\!$1474.4\pm6.0$\\
 $\Gamma_{f_0(1370)}$(MeV)&~$113\pm11$&~$108\pm11$\\
 $\phi_{980}$ & 0 (fixed)&0 (fixed)\\
 $\phi_{1370}$ &$241.5\pm6.3$&$181.7\pm8.4$\\
 $\phi_{\rm NR}$ &$217.0\pm3.7$&$232.2\pm3.7$\\
 $\phi_{1270}$, $\lambda=0$ & ~~\!$165\pm15$&~~\!$118\pm15$\\
  $\phi_{1270}$, $|\lambda|=1$ & 0 (fixed)&0 (fixed)\\ 
\hline
\end{tabular}\label{fitres2}
\end{center}
\end{table}

The mass and resonance parameters depend strongly on the final state in which they are measured, and the form of the
resonance fitting function. Thus we do not quote systematic errors on these values. 
The value found for the $f_0(980)$  mass in the Flatt\'e function $939.9\pm6.3$ MeV is lower than most determinations, although the observed peak value is close to 980 MeV, the estimated PDG value  \cite{PDG}. This is due to the interference from other resonances. The BES collaboration using the same functional form found a mass value of 965$\pm 8\pm$6 MeV in the $J/\psi\to \phi\pi^+\pi^-$ final state \cite{Ablikim:2004wn}. They also found roughly similar values of the coupling constants as ours,  $g_{\pi\pi}=165\pm 10 \pm 15$ MeV, and $g_{KK}/g_{\pi\pi}=4.21\pm 0.25\pm 0.21$. 
The PDG provides only estimated values for the $f_0(1370)$ mass of 1200$-$1500 MeV and width 200$-$500 MeV, respectively \cite{PDG}. Our result is within both of these ranges.

\subsection{Angular moments}

\def \t {\theta_{\pi\pi}}
\def \A {{\cal A}}

The angular moment distributions provide an additional way of
visualizing the effects of different resonances and their interferences, similar to a partial
wave analysis. This technique has been used in previous studies \cite{Lees:2012kx,*delAmoSanchez:2010yp}.

We define the angular moments 
$\langle Y_l^0\rangle$ as the efficiency corrected and background subtracted $\pi^+\pi^-$ invariant mass distributions, weighted by spherical harmonic functions
\begin{equation}
\langle Y_l^0 \rangle = \int_{-1}^{1}d\Gamma(m_{\pi\pi},\cos\t)Y_l^0(\cos\t)d\cos\t.
\end{equation}
The spherical harmonic functions satisfy
\begin{equation}
\int_{-1}^{1}Y_i^0(\cos\t)Y_j^0(\cos\t)d\cos\t=\frac{\delta_{ij}}{2\pi}.
\end{equation}

If we assume that no $\pi^+\pi^-$ partial-waves of a higher order than D-wave contribute, then we can express the differential decay rate ($d\Gamma$) derived from Eq. (3) in terms of S-, P-, and D-waves including helcity 0 and $\pm1$ components as
\begin{eqnarray}
d\Gamma(m_{\pi\pi},\cos\t)&=& 2\pi\left|\A_{S_0} Y_0^0(\cos \t)+\A_{P_0} e^{i\phi_{P_0}}Y_1^0(\cos \t)+\A_{D_0} e^{i\phi_{D_0}} Y_2^0(\cos \t)\right|^2\nonumber\\
&+&2\pi\left|\A_{P_{\pm 1}} e^{i\phi_{P_{\pm 1}}} \sqrt{\frac{3}{8\pi}}\sin\t +\A_{D_{\pm 1}}e^{i\phi_{D_{\pm 1}}} \sqrt{\frac{15}{8\pi}}\sin\t \cos\t\right|^2,
\end{eqnarray}
where $\A_{k_\lambda}$ and $\phi_{k_\lambda}$ are real-valued functions of $m_{\pi\pi}$, and we have factored out the S-wave phase. We then calculate the angular moments
\begin{eqnarray}
\sqrt{4\pi}\langle Y_0^0\rangle&=&\A_{S_0}^2+\A_{P_0}^2+\A_{D_0}^2+\A_{P_{\pm 1}}^2+\A_{D_{\pm 1}}^2,\nonumber\\
\sqrt{4\pi}\langle Y_1^0\rangle&=&2\A_{S_0}\A_{P_0}\cos\phi_{P_0}+\frac{4}{\sqrt{5}}\A_{P_0}\A_{D_0}\cos(\phi_{P_0}-\phi_{D_0})+8\sqrt{\frac{3}{5}}\A_{P_{\pm 1}}\A_{D_{\pm 1}}\cos(\phi_{P_{\pm 1}}-\phi_{D_{\pm 1}}),\nonumber\\
\sqrt{4\pi}\langle Y_2^0\rangle&=&\frac{2}{\sqrt{5}}\A_{P_0}^2+2\A_{S_0}\A_{D_0}\cos\phi_{D_0}+\frac{2\sqrt{5}}{7}\A_{D_0}^2-\frac{1}{\sqrt{5}}\A_{P_{\pm 1}}^2+\frac{\sqrt{5}}{7}\A_{D_{\pm 1}}^2,\nonumber\\
\sqrt{4\pi}\langle Y_3^0\rangle&=&6\sqrt{\frac{3}{35}}\A_{P_0}\A_{D_0}\cos(\phi_{P_0}-\phi_{D_0})+\frac{6}{\sqrt{35}}\A_{P_{\pm 1}}\A_{D_{\pm 1}}\cos(\phi_{P_{\pm 1}}-\phi_{D_{\pm 1}}),\nonumber\\
\sqrt{4\pi}\langle Y_4^0\rangle&=&\frac{6}{7}\A_{D_0}^2-\frac{4}{7}\A_{D_{\pm 1}}^2.
\end{eqnarray}

\begin{figure}[!htb]
\vskip -1cm
\begin{center}
    \includegraphics[width=6 in]{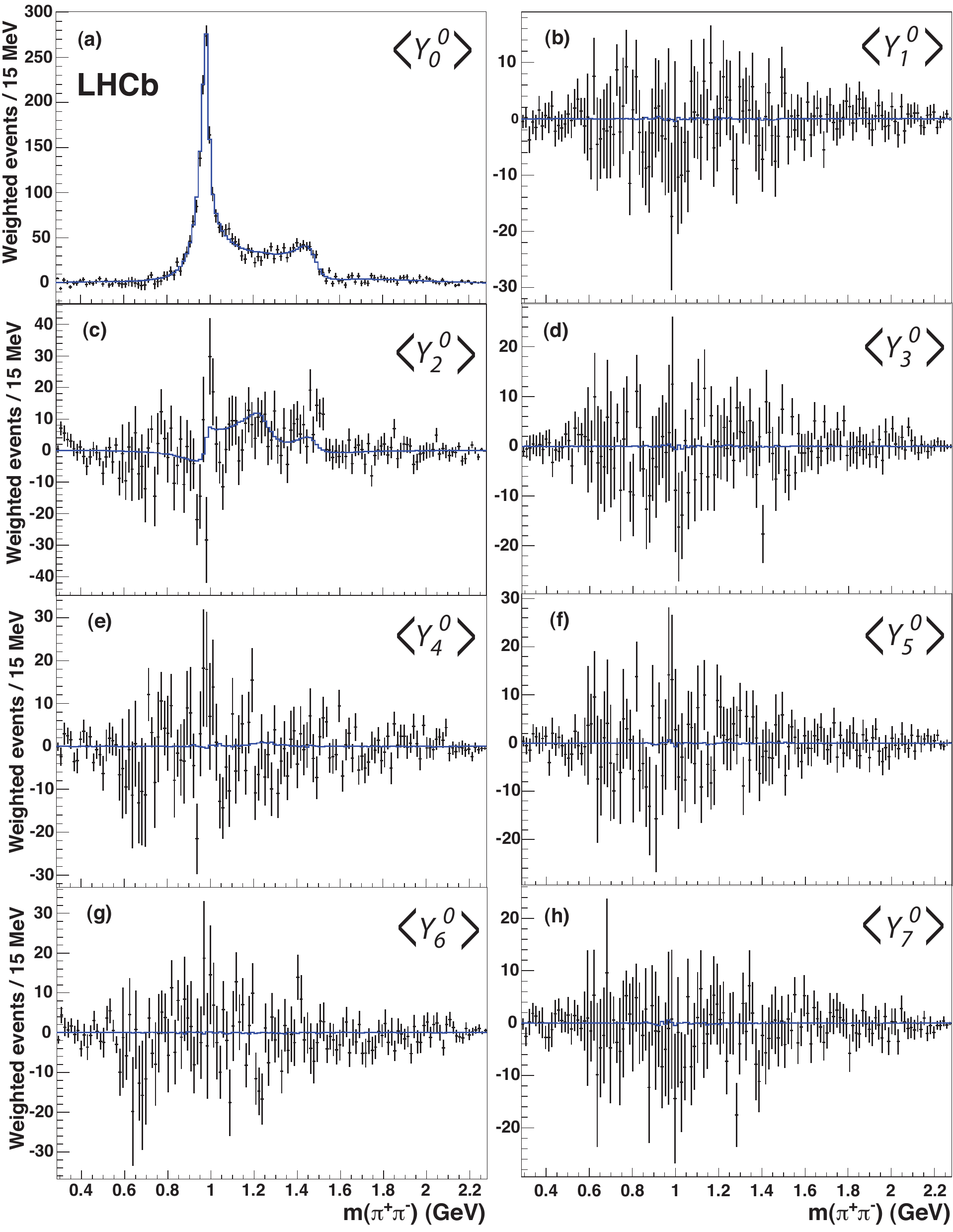}
\end{center}
\vskip -0.5cm
\caption{The $\pi^+\pi^-$ mass dependence of the spherical harmonic moments of $\cos \theta_{\pi\pi}$ after efficiency corrections and background subtraction:
(a) $\langle Y^0_0\rangle$,  (b) $\langle Y^0_1\rangle$, (c) $\langle Y^0_2\rangle$, (d) $\langle Y^0_3\rangle$, (e) $\langle Y^0_4\rangle$, (f) $\langle Y^0_5\rangle$, (g) $\langle Y^0_6\rangle$, and (h) $\langle Y^0_7\rangle$. 
The points with error bars are the data points and the solid curves are derived from the 3R+NR preferred model.}
\label{SPH2}
\end{figure}

Figure~\ref{SPH2} shows the distributions of the angular moments for the preferred solution.  In general the interpretation of these moments is that $\langle Y^0_0\rangle$ is the efficiency corrected and background subtracted event distribution, $\langle Y^0_1\rangle$  the interference of the sum of S-wave and P-wave and P-wave and D-wave amplitudes, $\langle Y^0_2\rangle$  the sum of the  P-wave, D-wave and the interference of S-wave and D-wave amplitudes, $\langle Y^0_3\rangle$  the interference between P-wave and D-wave, and $\langle Y^0_4\rangle$ the D-wave.  

In our data the $\langle Y^0_1\rangle$ distribution is consistent with zero, confirming the absence of any P-wave. We do observe the effects of the $f_2(1270)$ in the $ \langle Y^0_2\rangle$ distribution including the interferences with the S-waves. The other moments are consistent with the absence of any structure, as expected.

\section{Results}
\subsection{\boldmath \CP content}
The main result in this paper is that \CP-odd final states dominate. The $f_2(1270)$ helicity $\pm1$ yield is ($0.21\pm 0.65$)\%. As this represents a mixed \CP state, the upper limit on the \CP-even fraction due to this state is $<1.3$\,\% at 95\% confidence level (CL). Adding the $\rho(770)$ amplitude and repeating the fit shows that only an insignificant amount of $\rho(770)$ can be tolerated; in fact, the isospin violating $J/\psi \rho(770)$ final state is limited to $<$\,1.5\% at 95\% CL.  The sum of $f_2(1270)$ helicity $\pm1$ and $\rho(770)$ is limited to $<$\,2.3\% at 95\% CL.  In the $\pi^+\pi^-$ mass region within $\pm$90 MeV of 980 MeV, this limit improves to $<$\,0.6\% at 95\% CL.

\subsection{Total branching fraction ratio}
To avoid the uncertainties associated with absolute branching fraction measurements, we quote branching fractions relative to the $\Bsb\to J/\psi\phi$ channel.  The detection efficiency for this channel from Monte Carlo simulation is $(1.07\pm0.01)$\%, where the error is due to the limited Monte Carlo sample size.

\begin{figure}[t!]
\begin{center}
    \includegraphics[width=4.5in]{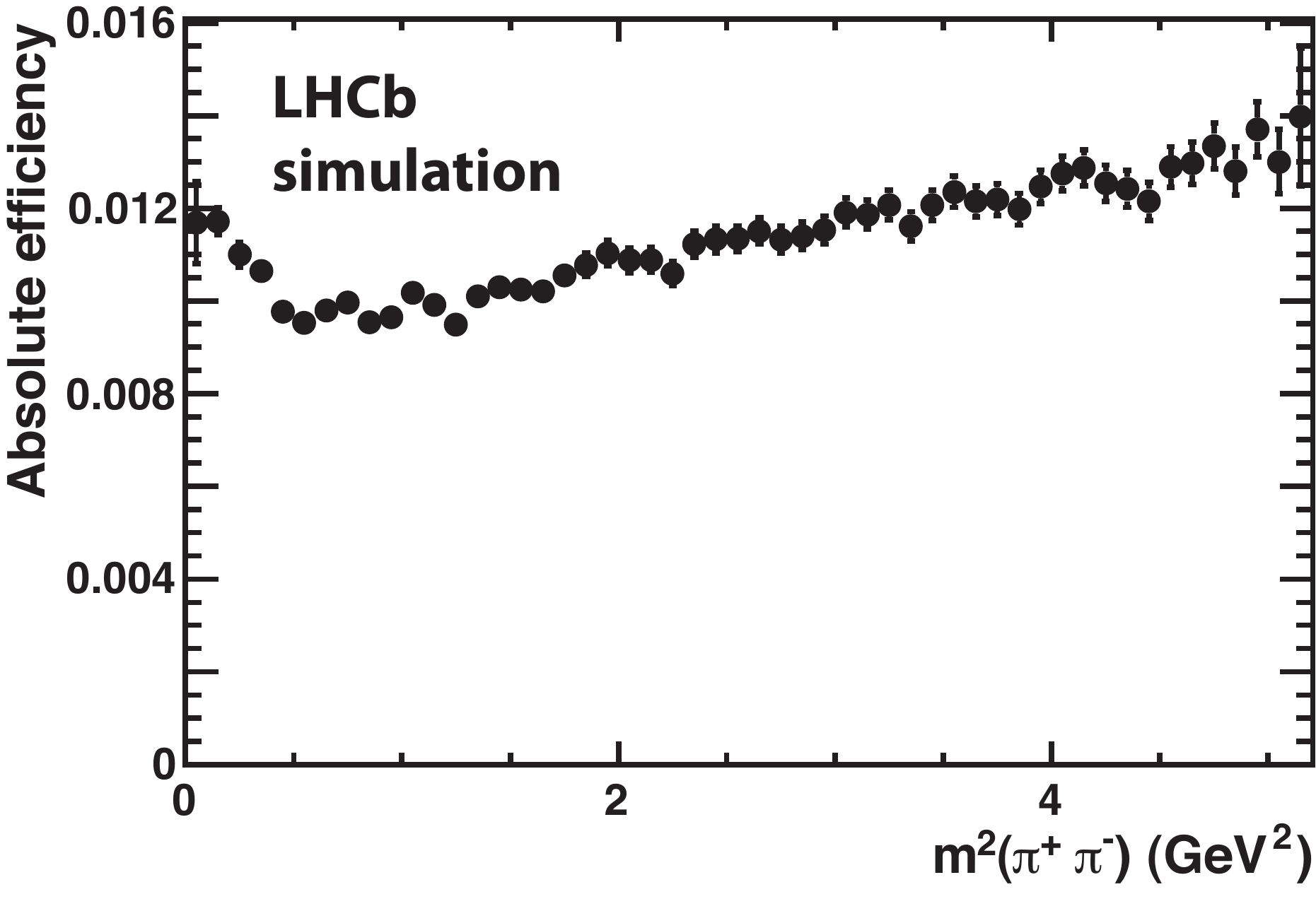}
    \caption{ Detection efficiency of $\Bsb\to J/\psi\pi^+\pi^-$ as a function of $s_{23}\equiv m^{2}(\pi^+\pi^-)$.}
\end{center}\label{Jpsipipieff}
\end{figure}

The simulated detection efficiency for $\Bsb\to J/\psi\pi^+\pi^-$ as a function of the $m^{2}(\pi^+\pi^-)$ is shown in Fig.~\ref{Jpsipipieff}. 
The simulation does not
 model the pion and kaon identification efficiencies with sufficient accuracy for our purposes. Therefore, we measure the
kaon identification efficiency with respect to the Monte Carlo simulation. We use samples of $D^{*+}\to \pi^+ D^0$, $D^0\to K^-\pi^+$ events selected without kaon identification to measure the kaon and pion efficiencies with respect to the simulation, and an additional sample of $K_s^0\to\pi^+\pi^-$ decay for pions. The identification  efficiency is measured in bins of \pt and $\eta$ and then the averages are weighted using the event distributions in the data. We find the correction to the $J/\psi\phi$ efficiency is 0.970 (two kaons) and that to the $J/\psi f_0$ efficiency is 0.973 (two pions). 
The additional correction due to particle identification then is 0.997$\pm$0.010. In addition, we re-weight the $\Bsb$ $p$ and $p_{\rm T}$ distributions in the simulation which lowers the $\pi^+\pi^-$  efficiency by 1.01\% with respect to the $K^+K^-$ efficiency.

Dividing the number of the $J/\psi\pi^+\pi^-$ signal events by the $J/\psi K^+K^-$ yield, applying the additional corrections as described above,
and taking into account ${\cal{B}}\left(\phi\to K^+K^-\right)=(48.9\pm 0.5)$\% \cite{PDG}, we find
\begin{equation}
\frac{{\cal{B}}\left(\Bsb\to J/\psi\pi^+\pi^-\right)}{{\cal{B}}\left(\Bsb\to J/\psi \phi\right)}
= (19.79\pm 0.47 \pm 0.52)\%.\nonumber
\end{equation}
Whenever two uncertainties are quoted the first is statistical and the second systematic. The latter will be discussed later in Section~\ref{sec:sys}.
This branching fraction ratio has not been previously measured.

\subsection{Relative resonance yields}
Next we evaluate the relative yields for the 3R+NR fit to the $J/\psi \pi^+\pi^-$ final state from the preferred solution.
We normalize the individual fit fractions reported in Table~\ref{ff3} by the sum.
These normalized  fit fractions are listed in Table~\ref{tab:nff} along with the branching fraction relative to $J/\psi\phi$, $\phi\to K^+K^-$,  defined as $R_r$, where $r$ refers to the particular final state under consideration. Thus
\begin{equation}
R_r=\frac{{\cal{B}}\left(\Bsb\to r\right)}{{\cal{B}}\left(\Bsb\to \jpsi\phi\right)}~.
\end{equation}
We use the difference between the preferred and alternate solutions found for the 3R+NR fit to assign a systematic uncertainty. Other systematic uncertainties are described in Section~\ref{sec:sys}.

The value found for $R_r$ for the $f_0(980)$, $0.139\pm0.006^{+0.025}_{-0.012}$, is consistent with the prediction of Ref.~\cite{Stone:2008ak}, and consistent with the our first observation
using 33\,pb$^{-1}$ of integrated luminosity \cite{Aaij:2011fx}, after multiplying by ${\cal{B}}\left(\phi\to K^+K^-\right)$.
The decay $\Bsb\to J/\psi f_0(1370)$ is now established. Previously both LHCb \cite{Aaij:2011fx} and Belle \cite{Li:2011pg} had seen evidence for this final state. The normalized $f_2(1270)$ helicity zero rate is 
(0.49$\pm$0.16)\% in the preferred model and (0.42$\pm$0.11)\% for the alternate solution. 

\begin{table}[h!t!p!]
\begin{center}
\caption{Normalized fit fractions (\%) for alternate and preferred 3R+NR models and the ratio $R$ (\%) relative to $\Bsb\to J/\psi\phi$. The numbers
for the $f_2(1270)$ refer only to the  $\lambda=0$ state.}
 \begin{tabular}{lccccc}
\hline
State & Preferred & Alternate& $R$ preferred  & $R$ alternate & Final $R$\\
\hline
$f_0(980)$    & $69.7\pm 2.3$  & $82.4\pm 2.3$ &$13.9\pm0.6$ &$16.3\pm0.6$             & $13.9\pm0.6^{+2.5}_{-1.2}$ \\
$f_0(1370)$   & $21.2\pm 2.7$ & $5.7\pm 0.7$  & $4.19\pm0.53$&$1.13\pm0.15$             & $4.19\pm0.53^{+0.12}_{-3.70}$\\
NR                  & $8.4\pm 1.5$& $11.3\pm 1.9$  & $1.66\pm0.31$  &$2.23\pm0.39$                        & $1.66\pm0.31^{+0.96}_{-0.08}$ \\
$f_2(1270)$    &$0.49\pm0.16$&$0.42\pm0.11$&$0.098\pm0.033$&$0.083\pm0.022$           &$0.098\pm0.033^{+0.006}_{-0.015}$\\
\hline
\end{tabular}\label{tab:nff}
\end{center}
\end{table}

\section{Systematic uncertainties}
\label{sec:sys}
Systematic uncertainties on the \CP-odd fraction are negligible.
In fact, using any of the alternate fits with different additional components does not introduce any significant fractions of \CP-odd final states.

The systematic uncertainties on the branching fraction ratios have several contributions listed in Table~\ref{tab:syserr}. Since $R_r$ is measured relative to $J/\psi\phi$ there is no systematic uncertainty due to differences in the tracking performance between data and simulation. The $J/\psi\phi$ P-wave yield is fully correlated with the S-wave yield whose uncertainty we estimate as 0.7\% by changing the signal PDF, and the background shape. By far the largest uncertainty in every rate, except the total, is caused by our choice of the preferred versus the alternate solutions. Using the difference between these fit results for the systematic uncertainty causes relatively large and asymmetric values. We also include systematic uncertainties due to the possible presence of the $\rho(770)$, the $f_0(1500)$, or the $f_0(600)$ resonances by taking the maximum difference between the fit including one of these resonances and our preferred solution, if the difference is larger than the one between the preferred and alternate 3R+NR fit. In the case of the $f_0(1500)$ the preferred solution is pathological in that it produces an unacceptably large $f_0(1370)$ component along with a 214\% component sum; therefore here we use the alternate solution that is much better behaved.

The uncertainty from Monte Carlo sample size for the mass dependent $\pi^+\pi^-$ efficiencies are accounted for in the statistical errors, a residual systematic uncertainty is included that results from allowed changes in the shape due to the distribution of the events. The size of these differences depends on the mass range for the particular component multiplied by the possible efficiency variation across this mass range. This is estimated as 1\% for the entire mass range and is smaller for individual resonances. Small uncertainties are introduced if the simulation does not
have the correct $\Bsb$ kinematic distributions. 
We are relatively insensitive to any these differences in the \Bsb $p$ and $p_{\rm T}$  distributions since we are measuring relative rates. These distributions are varied by changing the weights in each bin by plus and minus the statistical error in that bin.
 We see at most a 0.5\% change.
There is a 2\% systematic uncertainty assigned for the relative particle identification efficiencies. These efficiencies have been corrected from those predicted in the simulation by using pion data from $K_s^0\to\pi^+\pi^-$ decays and kaon and pion data from $D^{*\pm}\to \pi^{\pm}D^0(\overline{D}^0)$, $D^0(\overline{D}^0)\to K^{\mp}\pi^{\pm}$ decays. The uncertainty on the corrections is 0.5\% per track. The background modeling was changed by using a second-order polynomial shape in the $J/\psi\pi^+\pi^-$ mass fit giving a 0.6\% change in the signal yield. 
Since the  input $f_0(1370)$ mass and width parameters were allowed to vary within Gaussian constraints, there is no additional uncertainty to account for. 

\begin{table}[!htb]
\centering
\caption{Relative systematic uncertainties on $R$(\%).}
\label{tab:syserr}
\begin{tabular}{lcccccc}
\hline
Parameter &  Total  & $f_0(980)$ & $f_0(1370)$& NR&$f_2(1270)$, $\lambda=0$\\\hline
$m(\pi^+\pi^- )$ dependent effic.& 1.0 & 0.2 &0.2 &1.0&0.2\\
PID efficiency & 2.0& 2.0& 2.0& 2.0&2.0\\
$J/\psi\phi$ S-wave& 0.7&0.7&0.7&0.7&0.7\\
$\Bsb$ $p$ and $p_{\rm T}$ distributions &0.5& 0.5&0.5&0.5&0.5\\
Acceptance function&0 & 0.1 & 1.3& 1.4 & 3.9\\
${\cal{B}}\left(\phi\to K^+K^-\right)$& 1.0& 1.0& 1.0& 1.0&1.0\\
Background &0.6&0.6&0.6&0.6&0.6\\
Resonance fit &$-$  &$^{+18.2}_{-~8.0}$ &$^{+~0.8}_{-88.1}$ &$^{+57.6}_{-~3.7}$&$^{+~3.0}_{-15.8}$\\
\hline
Total & $\pm$2.7 &$^{+18.3}_{-~8.4}$ & $^{+~2.9}_{-88.2}$&$^{+57.7}_{-~4.8}$ &$^{+~5.5}_{-16.4}$ \\
\hline
\end{tabular}
\end{table}

The effect on the fit fractions of changing the acceptance function is also evaluated. Since the acceptance model was tested by its agreement with the $\Bsb\to J/\psi K^+K^-$ data in Fig.~\ref{cos_theta_Phi}, we vary the data so that the model does not
 fit as well. This is accomplished by increasing the minimum IP $\chi^2$ requirement from 9 to 12.25 on both of the kaon candidates, which has the effect of increasing the $\chi^2$/ndf of the fit to angular distributions by 1 unit. The Monte Carlo simulation of $\Bsb\to J/\psi \pi^+\pi^-$ with the changed requirement is then fitted to get an acceptance function. This acceptance function is then applied to the data with the original minimum IP $\chi^2$ cut of 9, and the likelihood fit is redone. 
The resulting fitted values from the preferred solution are compared with the original values in Table~\ref{tab:acceptsys}.
The changes are small and well within the statistical uncertainties.
\begin{table}[!htb]
\centering
\caption{Changes due to modified acceptance function.}
\label{tab:acceptsys}
\begin{tabular}{lccc}
\hline
Values  &	Original	& After change & Variation(\%)\\\hline
\multicolumn{4}{l}{Fit fractions}\\
$f_0(980)$& 	(107.1$\pm$3.5)\%	& 107.2\% &~0.1 \\
$f_2(1270)$~$\lambda=0$ &~\;\;(0.76$\pm$0.25)\% &~\;\;0.79\%&	3.9 \\
$f_2(1270)$~$|\lambda|=1$ &\;\;\;(0.33$\pm$1.00)\% &\;\;\;0.26\%&	21.2~ \\
$f_0(1370) $ & ~(32.6$\pm$4.1)\% &~31.2\% & 1.3\\
NR	&~(12.8$\pm$2.3)\%  &~12.7\% & 1.4 \\\hline
\multicolumn{4}{l}{$f_0(980)$ parameters}\\
$m_{f_0}$ (MeV)  &\!\!\!939.9$\pm$6.3 & 938.4& \!~~0.16\\
$g_{\pi\pi}$(MeV)   & 199$\pm$30 & 205 & 2.7\\
$g_{KK}/g_{\pi\pi}$ &~ 3.01$\pm$0.25  & 3.05& 1.3\\\hline
\multicolumn{4}{l}{$f_0(1370)$ parameters}\\
$m_{f_0}$ (MeV)  & \!\!\!1475.1$\pm$6.3&1476.4&0.09\\
$\Gamma$ (MeV)&~\,112.7$\pm$11.1&113.0&	0.27\\
\hline
\end{tabular}
\end{table}

\section{Conclusions}

We have studied the resonance structure of $\Bsb\rightarrow J/\psi \pi^+\pi^-$ using a modified Dalitz plot analysis where we also include the decay angle of the $J/\psi$. The decay distributions are formed from a series of final states described by individual $\pi^+\pi^-$ interfering decay amplitudes. The largest component is the $f_0(980)$ that is described by a Flatt\'e function. The data are best described by adding Breit-Wigner amplitudes for the $f_0(1370)$, the $f_2(1270)$  resonances and a non-resonance contribution.  
Adding a $\rho(770)$ into the fit does not improve the overall likelihood. Inclusion of  $f_0(600)$ or $f_0(1500)$ does not result in significant signals for these resonances.

Our three resonance plus non-resonance best fit is dominantly \CP-odd S-wave over the entire signal region. We also have a D-wave component arising from the $f_2(1270)$ resonance. Part of this corresponds to the $A_{20}$ amplitude which is also pure \CP-odd and is $(0.49\pm 0.16^{+0.02}_{-0.08})\%$ of the total rate. A mixed \CP part corresponding to the $A_{2\pm 1}$ amplitude is $(0.2\pm 0.7)\%$ of the total. 
Adding  this to the amount of allowed $\rho(770)$, less than $1.5$\% at 95\% CL, we find that the \CP-odd fraction is greater than 0.977 at 95\% CL.
Thus, the entire mass range can be used to study \CP violation with this almost pure \CP-odd final state.

The measured relative branching ratio is
\begin{equation}
\frac{{\cal{B}}\left(\Bsb\to J/\psi\pi^+\pi^-\right)}{{\cal{B}}\left(\Bsb\to J/\psi \phi\right)}
= (19.79\pm 0.47 \pm 0.52)\%, \nonumber
\end{equation}
where the first uncertainty is statistical and the second systematic.
The largest component is the $f_0(980)$ resonance. We also determine
\begin{equation}
\frac{{\cal{B}}\left(\Bsb\to J/\psi\pi^+\pi^-\right){\cal{B}}\left(f_0(980)\to \pi^+\pi^-\right)}{{\cal{B}}\left(\Bsb\to J/\psi \phi\right)}
= (13.9\pm0.6^{+2.5}_{-1.2})\%, \nonumber
\end{equation}
This state was predicted to exist and have a branching fraction about 10\% that of $J/\psi\phi$ \cite{Stone:2008ak}. Our new measurement is consistent with and somewhat larger than this prediction. Other models give somewhat higher rates
\cite{Colangelo:2010wg,*Colangelo:2010bg,*Fleischer:2011au,*ElBennich:2011gm,*Leitner:2010fq}. We also have firmly established the existence of the $J/\psi f_0(1370)$ final state in \Bsb decay.

\section*{Acknowledgements}

\noindent We express our gratitude to our colleagues in the CERN accelerator
departments for the excellent performance of the LHC. We thank the
technical and administrative staff at CERN and at the LHCb institutes,
and acknowledge support from the National Agencies: CAPES, CNPq,
FAPERJ and FINEP (Brazil); CERN; NSFC (China); CNRS/IN2P3 (France);
BMBF, DFG, HGF and MPG (Germany); SFI (Ireland); INFN (Italy); FOM and
NWO (The Netherlands); SCSR (Poland); ANCS (Romania); MinES of Russia and
Rosatom (Russia); MICINN, XuntaGal and GENCAT (Spain); SNSF and SER
(Switzerland); NAS Ukraine (Ukraine); STFC (United Kingdom); NSF
(USA). We also acknowledge the support received from the ERC under FP7
and the Region Auvergne.

\newpage

\ifx\mcitethebibliography\mciteundefinedmacro
\PackageError{LHCb.bst}{mciteplus.sty has not been loaded}
{This bibstyle requires the use of the mciteplus package.}\fi
\providecommand{\href}[2]{#2}


\begin{mcitethebibliography}{10}
\mciteSetBstSublistMode{n}
\mciteSetBstMaxWidthForm{subitem}{\alph{mcitesubitemcount})}
\mciteSetBstSublistLabelBeginEnd{\mcitemaxwidthsubitemform\space}
{\relax}{\relax}

\bibitem{Aaij:2011fx}
LHCb collaboration, R.~Aaij {\em et~al.},
  \ifthenelse{\boolean{articletitles}}{{\it {First observation of $B_s^0 \to
  J/\psi f_0(980)$ decays}},
  }{}\href{http://dx.doi.org/10.1016/j.physletb.2011.03.006}{Phys.\ Lett.\
  {\bf B698} (2011) 115}, \href{http://xxx.lanl.gov/abs/1102.0206}{{\tt
  arXiv:1102.0206}}\relax
\mciteBstWouldAddEndPuncttrue
\mciteSetBstMidEndSepPunct{\mcitedefaultmidpunct}
{\mcitedefaultendpunct}{\mcitedefaultseppunct}\relax
\EndOfBibitem
\bibitem{Li:2011pg}
Belle collaboration, J.~Li {\em et~al.},
  \ifthenelse{\boolean{articletitles}}{{\it {Observation of $B_s^0\to J/\psi
  f_0(980)$ and Evidence for $B_s^0\to J/\psi f_0(1370)$}},
  }{}\href{http://dx.doi.org/10.1103/PhysRevLett.106.121802}{Phys.\ Rev.\
  Lett.\  {\bf 106} (2011) 121802},
  \href{http://xxx.lanl.gov/abs/1102.2759}{{\tt arXiv:1102.2759}}\relax
\mciteBstWouldAddEndPuncttrue
\mciteSetBstMidEndSepPunct{\mcitedefaultmidpunct}
{\mcitedefaultendpunct}{\mcitedefaultseppunct}\relax
\EndOfBibitem
\bibitem{Abazov:2011hv}
D0 collaboration, V.~M. Abazov {\em et~al.},
  \ifthenelse{\boolean{articletitles}}{{\it {Measurement of the relative
  branching ratio of $B^0_s \to J/\psi f_{0}(980)$ to $B_{s}^{0} \to J/\psi
  \phi$}}, }{}\href{http://dx.doi.org/10.1103/PhysRevD.85.011103}{Phys.\ Rev.\
  {\bf D85} (2012) 011103}, \href{http://xxx.lanl.gov/abs/1110.4272}{{\tt
  arXiv:1110.4272}}\relax
\mciteBstWouldAddEndPuncttrue
\mciteSetBstMidEndSepPunct{\mcitedefaultmidpunct}
{\mcitedefaultendpunct}{\mcitedefaultseppunct}\relax
\EndOfBibitem
\bibitem{Aaltonen:2011nk}
CDF collaboration, T.~Aaltonen {\em et~al.},
  \ifthenelse{\boolean{articletitles}}{{\it {Measurement of branching ratio and
  $B_s^0$ lifetime in the decay $B_s^0 \rightarrow J/\psi f_0(980)$ at CDF}},
  }{}\href{http://dx.doi.org/10.1103/PhysRevD.84.052012}{Phys.\ Rev.\  {\bf
  D84} (2011) 052012}, \href{http://xxx.lanl.gov/abs/1106.3682}{{\tt
  arXiv:1106.3682}}\relax
\mciteBstWouldAddEndPuncttrue
\mciteSetBstMidEndSepPunct{\mcitedefaultmidpunct}
{\mcitedefaultendpunct}{\mcitedefaultseppunct}\relax
\EndOfBibitem
\bibitem{LHCb:2011ab}
LHCb collaboration, R.~Aaij {\em et~al.},
  \ifthenelse{\boolean{articletitles}}{{\it {Measurement of the \CP violating
  phase $\phi_s$ in $\Bsb\to J/\psi f_0(980)$}},
  }{}\href{http://dx.doi.org/10.1016/j.physletb.2012.01.017}{Phys.\ Lett.\
  {\bf B707} (2012) 497}, \href{http://xxx.lanl.gov/abs/1112.3056}{{\tt
  arXiv:1112.3056}}\relax
\mciteBstWouldAddEndPuncttrue
\mciteSetBstMidEndSepPunct{\mcitedefaultmidpunct}
{\mcitedefaultendpunct}{\mcitedefaultseppunct}\relax
\EndOfBibitem
\bibitem{LHCb:2011aa}
LHCb collaboration, R.~Aaij {\em et~al.},
  \ifthenelse{\boolean{articletitles}}{{\it {Measurement of the CP-violating
  phase $\phi_s$ in the decay $\Bs \to J/\psi \phi$}},
  }{}\href{http://dx.doi.org/10.1103/PhysRevLett..108.101803}{Phys.\ Rev.\
  Lett.\  {\bf 108} (2012) 101803},
  \href{http://xxx.lanl.gov/abs/1112.3183}{{\tt arXiv:1112.3183}}\relax
\mciteBstWouldAddEndPuncttrue
\mciteSetBstMidEndSepPunct{\mcitedefaultmidpunct}
{\mcitedefaultendpunct}{\mcitedefaultseppunct}\relax
\EndOfBibitem
\bibitem{CDF:2011af}
CDF collaboration, T.~Aaltonen {\em et~al.},
  \ifthenelse{\boolean{articletitles}}{{\it {Measurement of the CP-Violating
  phase $\beta_s$ in $B^0_s \to J/\psi \phi$ decays with the CDF II detector}},
  }{}\href{http://xxx.lanl.gov/abs/1112.1726}{{\tt arXiv:1112.1726}}\relax
\mciteBstWouldAddEndPuncttrue
\mciteSetBstMidEndSepPunct{\mcitedefaultmidpunct}
{\mcitedefaultendpunct}{\mcitedefaultseppunct}\relax
\EndOfBibitem
\bibitem{Abazov:2011ry}
D0 collaboration, V.~M. Abazov {\em et~al.},
  \ifthenelse{\boolean{articletitles}}{{\it {Measurement of the CP-violating
  phase $\phi_s^{J/\psi \phi}$ using the flavor-tagged decay $B_s^0 \rightarrow
  J/\psi \phi$ in 8 fb$^{-1}$ of $p \overline p$ collisions}},
  }{}\href{http://dx.doi.org/10.1103/PhysRevD.85.032006}{Phys.\ Rev.\  {\bf
  D85} (2012) 032006}, \href{http://xxx.lanl.gov/abs/1109.3166}{{\tt
  arXiv:1109.3166}}\relax
\mciteBstWouldAddEndPuncttrue
\mciteSetBstMidEndSepPunct{\mcitedefaultmidpunct}
{\mcitedefaultendpunct}{\mcitedefaultseppunct}\relax
\EndOfBibitem
\bibitem{Dalitz:1953cp}
R.~Dalitz, \ifthenelse{\boolean{articletitles}}{{\it {On the analysis of
  $\tau$-meson data and the nature of the $\tau$-meson}},
  }{}\href{http://dx.doi.org/10.1080/14786441008520365}{Phil.\ Mag.\  {\bf 44}
  (1953) 1068}\relax
\mciteBstWouldAddEndPuncttrue
\mciteSetBstMidEndSepPunct{\mcitedefaultmidpunct}
{\mcitedefaultendpunct}{\mcitedefaultseppunct}\relax
\EndOfBibitem
\bibitem{LHCb-det}
LHCb collaboration, A.~Alves~Jr. {\em et~al.},
  \ifthenelse{\boolean{articletitles}}{{\it {The LHCb detector at the LHC}},
  }{}\href{http://dx.doi.org/10.1088/1748-0221/3/08/S08005}{JINST {\bf 3}
  (2008) S08005}\relax
\mciteBstWouldAddEndPuncttrue
\mciteSetBstMidEndSepPunct{\mcitedefaultmidpunct}
{\mcitedefaultendpunct}{\mcitedefaultseppunct}\relax
\EndOfBibitem
\bibitem{PDG}
Particle Data Group, K.~Nakamura {\em et~al.},
  \ifthenelse{\boolean{articletitles}}{{\it {Review of particle physics}},
  }{}\href{http://dx.doi.org/10.1088/0954-3899/37/7A/075021}{J.\ Phys.\  {\bf
  G37} (2010) 075021}\relax
\mciteBstWouldAddEndPuncttrue
\mciteSetBstMidEndSepPunct{\mcitedefaultmidpunct}
{\mcitedefaultendpunct}{\mcitedefaultseppunct}\relax
\EndOfBibitem
\bibitem{Stone:2008ak}
S.~Stone and L.~Zhang, \ifthenelse{\boolean{articletitles}}{{\it {S-waves and
  the measurement of CP violating phases in $B_s$ Decays}},
  }{}\href{http://dx.doi.org/10.1103/PhysRevD.79.074024}{Phys.\ Rev.\  {\bf
  D79} (2009) 074024}, \href{http://xxx.lanl.gov/abs/0812.2832}{{\tt
  arXiv:0812.2832}}\relax
\mciteBstWouldAddEndPuncttrue
\mciteSetBstMidEndSepPunct{\mcitedefaultmidpunct}
{\mcitedefaultendpunct}{\mcitedefaultseppunct}\relax
\EndOfBibitem
\bibitem{Pivk:2004ty}
M.~Pivk and F.~R. Le~Diberder, \ifthenelse{\boolean{articletitles}}{{\it
  {${_SPlot}$: A Statistical tool to unfold data distributions}},
  }{}\href{http://dx.doi.org/10.1016/j.nima.2005.08.106}{Nucl.\ Instrum.\
  Meth.\  {\bf A555} (2005) 356},
  \href{http://xxx.lanl.gov/abs/physics/0402083}{{\tt
  arXiv:physics/0402083}}\relax
\mciteBstWouldAddEndPuncttrue
\mciteSetBstMidEndSepPunct{\mcitedefaultmidpunct}
{\mcitedefaultendpunct}{\mcitedefaultseppunct}\relax
\EndOfBibitem
\bibitem{Mizuk:2008me}
Belle collaboration, R.~Mizuk {\em et~al.},
  \ifthenelse{\boolean{articletitles}}{{\it {Observation of two resonance-like
  structures in the $\pi^+ \chi_{c1}$ mass distribution in exclusive
  $\overline{B}^0\to K^- \pi^+ \chi_{c1}$ decays}},
  }{}\href{http://dx.doi.org/10.1103/PhysRevD.78.072004}{Phys.\ Rev.\  {\bf
  D78} (2008) 072004}, \href{http://xxx.lanl.gov/abs/0806.4098}{{\tt
  arXiv:0806.4098}}\relax
\mciteBstWouldAddEndPuncttrue
\mciteSetBstMidEndSepPunct{\mcitedefaultmidpunct}
{\mcitedefaultendpunct}{\mcitedefaultseppunct}\relax
\EndOfBibitem
\bibitem{Z4430}
Belle collaboration, R.~Mizuk {\em et~al.},
  \ifthenelse{\boolean{articletitles}}{{\it {Dalitz analysis of $B \to K \pi^+
  \psi'$ decays and the $Z(4430)^+$}},
  }{}\href{http://dx.doi.org/10.1103/PhysRevD.80.031104}{Phys.\ Rev.\  {\bf
  D80} (2009) 031104}, \href{http://xxx.lanl.gov/abs/0905.2869}{{\tt
  arXiv:0905.2869}}\relax
\mciteBstWouldAddEndPuncttrue
\mciteSetBstMidEndSepPunct{\mcitedefaultmidpunct}
{\mcitedefaultendpunct}{\mcitedefaultseppunct}\relax
\EndOfBibitem
\bibitem{Blatt}
J.~M. Blatt and V.~F. Weisskopf, \ifthenelse{\boolean{articletitles}}{{\it
  {Theoretical Nuclear Physics}}, }{}Wiley/Springer-Verlag (1952)\relax
\mciteBstWouldAddEndPuncttrue
\mciteSetBstMidEndSepPunct{\mcitedefaultmidpunct}
{\mcitedefaultendpunct}{\mcitedefaultseppunct}\relax
\EndOfBibitem
\bibitem{Flatte:1976xv}
S.~M. Flatt\'e, \ifthenelse{\boolean{articletitles}}{{\it {On the Nature of 0+
  Mesons}}, }{}\href{http://dx.doi.org/10.1016/0370-2693(76)90655-9}{Phys.\
  Lett.\  {\bf B63} (1976) 228}\relax
\mciteBstWouldAddEndPuncttrue
\mciteSetBstMidEndSepPunct{\mcitedefaultmidpunct}
{\mcitedefaultendpunct}{\mcitedefaultseppunct}\relax
\EndOfBibitem
\bibitem{Sjostrand:2006za}
T.~Sj{\"o}strand, S.~Mrenna, and P.~Skands,
  \ifthenelse{\boolean{articletitles}}{{\it {PYTHIA 6.4 Physics and Manual}},
  }{}\href{http://dx.doi.org/10.1088/1126-6708/2006/05/026}{JHEP {\bf 0605}
  (2006) 026}, \href{http://xxx.lanl.gov/abs/hep-ph/0603175}{{\tt
  arXiv:hep-ph/0603175}}\relax
\mciteBstWouldAddEndPuncttrue
\mciteSetBstMidEndSepPunct{\mcitedefaultmidpunct}
{\mcitedefaultendpunct}{\mcitedefaultseppunct}\relax
\EndOfBibitem
\bibitem{LHCb-PROC-2011-005}
I.~Belyaev {\em et~al.}, \ifthenelse{\boolean{articletitles}}{{\it {Handling of
  the generation of primary events in \gauss, the \lhcb simulation framework}},
  }{}\href{http://dx.doi.org/10.1109/NSSMIC.2010.5873949}{Nuclear Science
  Symposium Conference Record (NSS/MIC) {\bf IEEE} (2010) 1155}\relax
\mciteBstWouldAddEndPuncttrue
\mciteSetBstMidEndSepPunct{\mcitedefaultmidpunct}
{\mcitedefaultendpunct}{\mcitedefaultseppunct}\relax
\EndOfBibitem
\bibitem{Agostinelli:2002hh}
GEANT4 collaboration, S.~Agostinelli {\em et~al.},
  \ifthenelse{\boolean{articletitles}}{{\it {GEANT4: A Simulation toolkit}},
  }{}\href{http://dx.doi.org/10.1016/S0168-9002(03)01368-8}{Nucl.\ Instrum.\
  Meth.\  {\bf A506} (2003) 250}\relax
\mciteBstWouldAddEndPuncttrue
\mciteSetBstMidEndSepPunct{\mcitedefaultmidpunct}
{\mcitedefaultendpunct}{\mcitedefaultseppunct}\relax
\EndOfBibitem
\bibitem{LHCb-PROC-2011-006}
M.~Clemencic {\em et~al.}, \ifthenelse{\boolean{articletitles}}{{\it {The \lhcb
  Simulation Application, Gauss: Design, Evolution and Experience}},
  }{}\href{http://dx.doi.org/10.1088/1742-6596/331/3/032023}{{Journal of
  Physics: Conference Series} {\bf 331} (2011), no.~3 032023}\relax
\mciteBstWouldAddEndPuncttrue
\mciteSetBstMidEndSepPunct{\mcitedefaultmidpunct}
{\mcitedefaultendpunct}{\mcitedefaultseppunct}\relax
\EndOfBibitem
\bibitem{Allison:1993dn}
J.~Allison, \ifthenelse{\boolean{articletitles}}{{\it {Multiquadric radial
  basis functions for representing multidimensional high-energy physics data}},
  }{}\href{http://dx.doi.org/10.1016/0010-4655(93)90184-E}{Comput.\ Phys.\
  Commun.\  {\bf 77} (1993) 377}\relax
\mciteBstWouldAddEndPuncttrue
\mciteSetBstMidEndSepPunct{\mcitedefaultmidpunct}
{\mcitedefaultendpunct}{\mcitedefaultseppunct}\relax
\EndOfBibitem
\bibitem{Muramatsu:2002jp}
CLEO collaboration, H.~Muramatsu {\em et~al.},
  \ifthenelse{\boolean{articletitles}}{{\it {Dalitz analysis of $D^0 \to K^0_S
  \pi^+\pi^-$}}, }{}\href{http://dx.doi.org/10.1103/PhysRevLett.89.251802,
  10.1103/PhysRevLett.89.251802}{Phys.\ Rev.\ Lett.\  {\bf 89} (2002) 251802},
  \href{http://xxx.lanl.gov/abs/hep-ex/0207067}{{\tt
  arXiv:hep-ex/0207067}}\relax
\mciteBstWouldAddEndPuncttrue
\mciteSetBstMidEndSepPunct{\mcitedefaultmidpunct}
{\mcitedefaultendpunct}{\mcitedefaultseppunct}\relax
\EndOfBibitem
\bibitem{Aitala:2000xt}
E791 collaboration, E.~M. Aitala {\em et~al.},
  \ifthenelse{\boolean{articletitles}}{{\it {Study of the $D^+_{s}\to \pi^-
  \pi^+ \pi^+$ decay and measurement of $f_0$ masses and widths}},
  }{}\href{http://dx.doi.org/10.1103/PhysRevLett.86.765}{Phys.\ Rev.\ Lett.\
  {\bf 86} (2001) 765}, \href{http://xxx.lanl.gov/abs/hep-ex/0007027}{{\tt
  arXiv:hep-ex/0007027}}\relax
\mciteBstWouldAddEndPuncttrue
\mciteSetBstMidEndSepPunct{\mcitedefaultmidpunct}
{\mcitedefaultendpunct}{\mcitedefaultseppunct}\relax
\EndOfBibitem
\bibitem{Baker:1983tu}
S.~Baker and R.~D. Cousins, \ifthenelse{\boolean{articletitles}}{{\it
  {Clarification of the use of $\chi^2$ and likelihood functions in fits to
  histograms}}, }{}\href{http://dx.doi.org/10.1016/0167-5087(84)90016-4}{Nucl.\
  Instrum.\ Meth.\  {\bf 221} (1984) 437}\relax
\mciteBstWouldAddEndPuncttrue
\mciteSetBstMidEndSepPunct{\mcitedefaultmidpunct}
{\mcitedefaultendpunct}{\mcitedefaultseppunct}\relax
\EndOfBibitem
\bibitem{Ablikim:2004wn}
BES collaboration, M.~Ablikim {\em et~al.},
  \ifthenelse{\boolean{articletitles}}{{\it {Resonances in $J/\psi \to \phi
  \pi^+\pi^-$ and $\phi K^+ K^-$}},
  }{}\href{http://dx.doi.org/10.1016/j.physletb.2004.12.041}{Phys.\ Lett.\
  {\bf B607} (2005) 243}, \href{http://xxx.lanl.gov/abs/hep-ex/0411001}{{\tt
  arXiv:hep-ex/0411001}}\relax
\mciteBstWouldAddEndPuncttrue
\mciteSetBstMidEndSepPunct{\mcitedefaultmidpunct}
{\mcitedefaultendpunct}{\mcitedefaultseppunct}\relax
\EndOfBibitem
\bibitem{Lees:2012kx}
BABAR Collaboration, J.~Lees, \ifthenelse{\boolean{articletitles}}{{\it {Study
  of CP violation in Dalitz-plot analyses of $\Bz \to K^+K^-K^0_S$, $B^+ \to
  K^+K^-K^+$, and $B^+ \to K^0_SK^0_SK^+$}},
  }{}\href{http://xxx.lanl.gov/abs/1201.5897}{{\tt arXiv:1201.5897}}\relax
\mciteBstWouldAddEndPuncttrue
\mciteSetBstMidEndSepPunct{\mcitedefaultmidpunct}
{\mcitedefaultendpunct}{\mcitedefaultseppunct}\relax
\EndOfBibitem
\bibitem{delAmoSanchez:2010yp}
BABAR Collaboration, P.~del Amo~Sanchez {\em et~al.},
  \ifthenelse{\boolean{articletitles}}{{\it {Dalitz plot analysis of $D_s^+ \to
  K^+ K^- \pi^+$}}, }{}Phys.\ Rev.\  {\bf D83} (2011) 052001,
  \href{http://xxx.lanl.gov/abs/1011.4190}{{\tt arXiv:1011.4190}}\relax
\mciteBstWouldAddEndPuncttrue
\mciteSetBstMidEndSepPunct{\mcitedefaultmidpunct}
{\mcitedefaultendpunct}{\mcitedefaultseppunct}\relax
\EndOfBibitem
\bibitem{Colangelo:2010wg}
P.~Colangelo, F.~De~Fazio, and W.~Wang,
  \ifthenelse{\boolean{articletitles}}{{\it {Nonleptonic $B_s$ to charmonium
  decays: analyses in pursuit of determining the weak phase $\beta_s$}},
  }{}\href{http://dx.doi.org/10.1103/PhysRevD.83.094027}{Phys.\ Rev.\  {\bf
  D83} (2011) 094027}, \href{http://xxx.lanl.gov/abs/1009.4612}{{\tt
  arXiv:1009.4612}}\relax
\mciteBstWouldAddEndPuncttrue
\mciteSetBstMidEndSepPunct{\mcitedefaultmidpunct}
{\mcitedefaultendpunct}{\mcitedefaultseppunct}\relax
\EndOfBibitem
\bibitem{Colangelo:2010bg}
P.~Colangelo, F.~De~Fazio, and W.~Wang,
  \ifthenelse{\boolean{articletitles}}{{\it {$B_s\to f_0(980)$ form factors and
  $B_s$ decays into $f_0(980)$}},
  }{}\href{http://dx.doi.org/10.1103/PhysRevD.81.074001}{Phys.\ Rev.\  {\bf
  D81} (2010) 074001}, \href{http://xxx.lanl.gov/abs/1002.2880}{{\tt
  arXiv:1002.2880}}\relax
\mciteBstWouldAddEndPuncttrue
\mciteSetBstMidEndSepPunct{\mcitedefaultmidpunct}
{\mcitedefaultendpunct}{\mcitedefaultseppunct}\relax
\EndOfBibitem
\bibitem{Fleischer:2011au}
R.~Fleischer, R.~Knegjens, and G.~Ricciardi,
  \ifthenelse{\boolean{articletitles}}{{\it {Anatomy of $B^0_{s,d} \to J/\psi
  f_0(980)$}}, }{}\href{http://xxx.lanl.gov/abs/1109.1112}{{\tt
  arXiv:1109.1112}}\relax
\mciteBstWouldAddEndPuncttrue
\mciteSetBstMidEndSepPunct{\mcitedefaultmidpunct}
{\mcitedefaultendpunct}{\mcitedefaultseppunct}\relax
\EndOfBibitem
\bibitem{ElBennich:2011gm}
B.~El-Bennich, J.~de~Melo, O.~Leitner, B.~Loiseau, and J.~Dedonder,
  \ifthenelse{\boolean{articletitles}}{{\it {New physics in $\Bs\to J/\psi\phi$
  decays?}}, }{}\href{http://xxx.lanl.gov/abs/1111.6955}{{\tt
  arXiv:1111.6955}}\relax
\mciteBstWouldAddEndPuncttrue
\mciteSetBstMidEndSepPunct{\mcitedefaultmidpunct}
{\mcitedefaultendpunct}{\mcitedefaultseppunct}\relax
\EndOfBibitem
\bibitem{Leitner:2010fq}
O.~Leitner, J.-P. Dedonder, B.~Loiseau, and B.~El-Bennich,
  \ifthenelse{\boolean{articletitles}}{{\it {Scalar resonance effects on the
  $\Bs-\Bsb$ mixing angle}},
  }{}\href{http://dx.doi.org/10.1103/PhysRevD.82.076006}{Phys.\ Rev.\  {\bf
  D82} (2010) 076006}, \href{http://xxx.lanl.gov/abs/1003.5980}{{\tt
  arXiv:1003.5980}}\relax
\mciteBstWouldAddEndPuncttrue
\mciteSetBstMidEndSepPunct{\mcitedefaultmidpunct}
{\mcitedefaultendpunct}{\mcitedefaultseppunct}\relax
\EndOfBibitem
\end{mcitethebibliography}

\end{document}